\documentclass[11pt,twocolumn,tighten]{aastex63}
\usepackage{hyperref}
\usepackage{color}
\usepackage{url}
\usepackage{natbib}
\usepackage{graphicx}
\usepackage{xfrac}
\usepackage{ulem}
\usepackage{CJK}
\usepackage{threeparttable}
\usepackage{multirow}
\usepackage{amsmath}
\begin{document}
\title{Unveiling Hidden Lyman Alpha Emitters in the DESI DR1 Data}
\correspondingauthor{Jui-Kuan Chan, Ting-Wen Lan}
\email{jkchan@ntu.edu.tw, twlan@ntu.edu.tw}

\author[0009-0009-4339-0102]{Jui-Kuan Chan}
\affiliation{Department of Physics and Center for Theoretical Physics, National Taiwan University, No. 1, Sec. 4, Roosevelt Rd., Taipei 10617, Taiwan}
\affiliation{Graduate Institute of Astrophysics, National Taiwan University, No. 1, Sec. 4, Roosevelt Rd., Taipei 10617, Taiwan}
\author[0000-0001-8857-7020]{Ting-Wen Lan}
\affiliation{Graduate Institute of Astrophysics, National Taiwan University, No. 1, Sec. 4, Roosevelt Rd., Taipei 10617, Taiwan}
\affiliation{Department of Physics and Center for Theoretical Physics, National Taiwan University, No. 1, Sec. 4, Roosevelt Rd., Taipei 10617, Taiwan}
\affiliation{Institute of Astronomy and Astrophysics, Academia Sinica, No. 1, Sec. 4, Roosevelt Rd., Taipei 10617, Taiwan}
\author[0000-0002-7738-6875]{J. Xavier Prochaska}
\affiliation{Department of Astronomy and Astrophysics, University of California at Santa Cruz, 1156 High Street, Santa Cruz, CA 95064, USA}
\affiliation{University of California Observatories, Lick Observatory, 1156 High Street, Santa Cruz, CA 95064, USA}
\affiliation{Kavli Institute for the Physics and Mathematics of the Universe, University of Tokyo, Kashiwa 277-8583, Japan}
\affiliation{Division of Science, National Astronomical Observatory of Japan,2-21-1 Osawa, Mitaka, Tokyo 181-8588, Japan}
\author[0000-0002-6186-5476]{Shun Saito}
\affiliation{Institute for Multi-messenger Astrophysics and Cosmology, Department of Physics, Missouri University of Science and Technology,
1315 N Pine Street, Rolla, MO 65409, USA}
\affiliation{Kavli Institute for the Physics and Mathematics of the Universe, University of Tokyo, Kashiwa 277-8583, Japan}

\author{J.~Aguilar}
\affiliation{Lawrence Berkeley National Laboratory, 1 Cyclotron Road, Berkeley, CA 94720, USA}
\author[0000-0001-6098-7247]{S.~Ahlen}
\affiliation{Department of Physics, Boston University, 590 Commonwealth Avenue, Boston, MA 02215 USA}
\author[0000-0001-9712-0006]{D.~Bianchi}
\affiliation{Dipartimento di Fisica ``Aldo Pontremoli'', Universit\`a degli Studi di Milano, Via Celoria 16, I-20133 Milano, Italy}
\affiliation{INAF-Osservatorio Astronomico di Brera, Via Brera 28, 20122 Milano, Italy}
\author{D.~Brooks}
\affiliation{Department of Physics \& Astronomy, University College London, Gower Street, London, WC1E 6BT, UK}
\author[0000-0002-2169-0595]{A.~Cuceu}
\affiliation{Lawrence Berkeley National Laboratory, 1 Cyclotron Road, Berkeley, CA 94720, USA}
\author[0000-0002-1769-1640]{A.~de la Macorra}
\affiliation{Instituto de F\'{\i}sica, Universidad Nacional Aut\'{o}noma de M\'{e}xico,  Circuito de la Investigaci\'{o}n Cient\'{\i}fica, Ciudad Universitaria, Cd. de M\'{e}xico  C.~P.~04510,  M\'{e}xico}
\author[0000-0002-5665-7912]{Biprateep~Dey}
\affiliation{Department of Astronomy \& Astrophysics, University of Toronto, Toronto, ON M5S 3H4, Canada}
\affiliation{Department of Physics \& Astronomy and Pittsburgh Particle Physics, Astrophysics, and Cosmology Center (PITT PACC), University of Pittsburgh, 3941 O'Hara Street, Pittsburgh, PA 15260, USA}
\author{P.~Doel}
\affiliation{Department of Physics \& Astronomy, University College London, Gower Street, London, WC1E 6BT, UK}
\author[0000-0002-3033-7312]{A.~Font-Ribera}
\affiliation{Instituci\'{o} Catalana de Recerca i Estudis Avan\c{c}ats, Passeig de Llu\'{\i}s Companys, 23, 08010 Barcelona, Spain}
\affiliation{Institut de F\'{i}sica d’Altes Energies (IFAE), The Barcelona Institute of Science and Technology, Edifici Cn, Campus UAB, 08193, Bellaterra (Barcelona), Spain}
\author[0000-0002-2890-3725]{J.~E.~Forero-Romero}
\affiliation{Departamento de F\'isica, Universidad de los Andes, Cra. 1 No. 18A-10, Edificio Ip, CP 111711, Bogot\'a, Colombia}
\affiliation{Observatorio Astron\'omico, Universidad de los Andes, Cra. 1 No. 18A-10, Edificio H, CP 111711 Bogot\'a, Colombia}
\author[0000-0001-9632-0815]{E.~Gaztañaga}
\affiliation{Institut d'Estudis Espacials de Catalunya (IEEC), c/ Esteve Terradas 1, Edifici RDIT, Campus PMT-UPC, 08860 Castelldefels, Spain}
\affiliation{Institute of Cosmology and Gravitation, University of Portsmouth, Dennis Sciama Building, Portsmouth, PO1 3FX, UK}
\affiliation{Institute of Space Sciences, ICE-CSIC, Campus UAB, Carrer de Can Magrans s/n, 08913 Bellaterra, Barcelona, Spain}
\author[0000-0003-3142-233X]{Satya~{Gontcho A Gontcho}}
\affiliation{University of Virginia, Department of Astronomy, Charlottesville, VA 22904, USA}
\author{G.~Gutierrez}
\affiliation{Fermi National Accelerator Laboratory, PO Box 500, Batavia, IL 60510, USA}
\author[0000-0003-1197-0902]{C.~Hahn}
\affiliation{Department of Astronomy, University of Texas at Austin, 2515 Speedway, TX 78712, USA}
\author[0000-0001-8528-3473]{J.~Jimenez}
\affiliation{Institut de F\'{i}sica d’Altes Energies (IFAE), The Barcelona Institute of Science and Technology, Edifici Cn, Campus UAB, 08193, Bellaterra (Barcelona), Spain}
\author[0000-0003-0201-5241]{R.~Joyce}
\affiliation{NSF NOIRLab, 950 N. Cherry Ave., Tucson, AZ 85719, USA}
\author[0000-0002-0000-2394]{S.~Juneau}
\affiliation{NSF NOIRLab, 950 N. Cherry Ave., Tucson, AZ 85719, USA}
\author[0000-0002-8828-5463]{D.~Kirkby}
\affiliation{Department of Physics and Astronomy, University of California, Irvine, 92697, USA}
\author[0000-0001-6356-7424]{A.~Kremin}
\affiliation{Lawrence Berkeley National Laboratory, 1 Cyclotron Road, Berkeley, CA 94720, USA}
\author[0000-0003-1838-8528]{M.~Landriau}
\affiliation{Lawrence Berkeley National Laboratory, 1 Cyclotron Road, Berkeley, CA 94720, USA}
\author[0000-0003-4962-8934]{M.~Manera}
\affiliation{Departament de F\'{i}sica, Serra H\'{u}nter, Universitat Aut\`{o}noma de Barcelona, 08193 Bellaterra (Barcelona), Spain}
\affiliation{Institut de F\'{i}sica d’Altes Energies (IFAE), The Barcelona Institute of Science and Technology, Edifici Cn, Campus UAB, 08193, Bellaterra (Barcelona), Spain}
\author[0000-0002-1125-7384]{A.~Meisner}
\affiliation{NSF NOIRLab, 950 N. Cherry Ave., Tucson, AZ 85719, USA}
\author{R.~Miquel}
\affiliation{Instituci\'{o} Catalana de Recerca i Estudis Avan\c{c}ats, Passeig de Llu\'{\i}s Companys, 23, 08010 Barcelona, Spain}
\affiliation{Institut de F\'{i}sica d’Altes Energies (IFAE), The Barcelona Institute of Science and Technology, Edifici Cn, Campus UAB, 08193, Bellaterra (Barcelona), Spain}
\author[0000-0002-2733-4559]{J.~Moustakas}
\affiliation{Department of Physics and Astronomy, Siena University, 515 Loudon Road, Loudonville, NY 12211, USA}
\author[0000-0001-9070-3102]{S.~Nadathur}
\affiliation{Institute of Cosmology and Gravitation, University of Portsmouth, Dennis Sciama Building, Portsmouth, PO1 3FX, UK}
\author[0000-0002-0644-5727]{W.~J.~Percival}
\affiliation{Department of Physics and Astronomy, University of Waterloo, 200 University Ave W, Waterloo, ON N2L 3G1, Canada}
\affiliation{Perimeter Institute for Theoretical Physics, 31 Caroline St. North, Waterloo, ON N2L 2Y5, Canada}
\affiliation{Waterloo Centre for Astrophysics, University of Waterloo, 200 University Ave W, Waterloo, ON N2L 3G1, Canada}
\author{C.~Poppett}
\affiliation{Lawrence Berkeley National Laboratory, 1 Cyclotron Road, Berkeley, CA 94720, USA}
\affiliation{Space Sciences Laboratory, University of California, Berkeley, 7 Gauss Way, Berkeley, CA  94720, USA}
\affiliation{University of California, Berkeley, 110 Sproul Hall \#5800 Berkeley, CA 94720, USA}
\author[0000-0001-7145-8674]{F.~Prada}
\affiliation{Instituto de Astrof\'{i}sica de Andaluc\'{i}a (CSIC), Glorieta de la Astronom\'{i}a, s/n, E-18008 Granada, Spain}
\author[0000-0001-6979-0125]{I.~P\'erez-R\`afols}
\affiliation{Departament de F\'isica, EEBE, Universitat Polit\`ecnica de Catalunya, c/Eduard Maristany 10, 08930 Barcelona, Spain}
\author{G.~Rossi}
\affiliation{Department of Physics and Astronomy, Sejong University, 209 Neungdong-ro, Gwangjin-gu, Seoul 05006, Republic of Korea}
\author[0000-0002-9646-8198]{E.~Sanchez}
\affiliation{CIEMAT, Avenida Complutense 40, E-28040 Madrid, Spain}
\author[0000-0002-5042-5088]{D.~Schlegel}
\affiliation{Lawrence Berkeley National Laboratory, 1 Cyclotron Road, Berkeley, CA 94720, USA}
\author{M.~Schubnell}
\affiliation{Department of Physics, University of Michigan, 450 Church Street, Ann Arbor, MI 48109, USA}
\affiliation{University of Michigan, 500 S. State Street, Ann Arbor, MI 48109, USA}
\author[0000-0002-3461-0320]{J.~Silber}
\affiliation{Lawrence Berkeley National Laboratory, 1 Cyclotron Road, Berkeley, CA 94720, USA}
\author{D.~Sprayberry}
\affiliation{NSF NOIRLab, 950 N. Cherry Ave., Tucson, AZ 85719, USA}
\author[0000-0003-1704-0781]{G.~Tarl\'{e}}
\affiliation{University of Michigan, 500 S. State Street, Ann Arbor, MI 48109, USA}
\author{B.~A.~Weaver}
\affiliation{NSF NOIRLab, 950 N. Cherry Ave., Tucson, AZ 85719, USA}
\author[0000-0002-6684-3997]{H.~Zou}
\affiliation{National Astronomical Observatories, Chinese Academy of Sciences, A20 Datun Road, Chaoyang District, Beijing, 100101, P.~R.~China}

\begin{abstract}

We present an automatic method based on machine-learning convolutional neural network (CNN) architecture to detect Lyman alpha emitters (LAE) hidden in the Data Release 1 spectroscopic dataset of the Dark Energy Spectroscopic Instrument (DESI). Those LAEs mostly have incorrect redshift estimations because the current DESI pipeline is not designed to detect and measure the redshifts of galaxies at $z>2$. To uncover those sources, we first visually inspect 4,134 DESI spectra and construct a sample, consisting of both LAEs and non-LAEs, for training and testing the CNN-based model to (1) detect LAEs in DESI spectra and (2) determine their Ly$\alpha$ redshifts. The final model yields $95.2\%$ purity and $95.9\%$ completeness for detecting LAEs.
We apply this model to approximately $2\times10^{6}$ spectra of sources targeted as emission-line galaxies and detect 19,685 LAEs from $z\sim2$ to $3.5$ within 12 minutes with a single GPU, illustrating the high efficiency of this model for identifying LAEs. The detected LAEs are mostly at the bright end of the luminosity function with Ly$\alpha$ luminosity $L_{\rm Ly\alpha} \gtrsim 10^{43}$ erg/s. The high signal-to-noise composite spectrum of the detected LAEs further shows various spectral features, including P-Cygni profiles of metal lines and Mg~\textsc{ii} emission lines, possible indicators of Lyman continuum escape fraction, revealing the rich astrophysical information in this LAE sample. 
Finally, this sample can be used to train and validate the pipelines for redshift determination of LAEs for the preparation of the DESI-II survey.
\end{abstract}

\section{Introduction}

Lyman-alpha emitters (LAEs) are a galaxy population that produces strong Lyman-alpha emission lines \citep[e.g., equivalent width $> 20 \, \mathrm{\AA}$;][]{Ouchi_LAE_2020}, powered by either star-formation or active galactic nuclei (AGNs) \citep[see][for a review]{Ouchi_LAE_2020}.  
LAEs have been considered to be a crucial galaxy population in the context of galaxy evolution \citep[e.g.,][]{Ouchi_LAE_2020, Firestone_2025_SFH, Nagaraj_2025_LF} as well as a potential population that contributes to the reionization of the Universe \citep[e.g.,][]{2020ApJ...889..161N}. Moreover, the strong asymmetric emission line profile of LAEs has been used as a redshift indicator at $z>2$ observed in optical wavelengths, making LAEs as one of the key tracers of large-scale structure for current and future cosmological surveys, such as the Hobby-Eberly Telescope Dark Energy Experiment \citep[HETDEX,][]{Hill_HETDEX_2008, HETDEX_2021}, the next stage of Dark Energy Spectroscopic Instrument \citep[DESI-II,][]{Schlegel_desi2_2022}, MegaMapper \citep{Schlegel_megamapper_2022}, and Spec-S5 \citep{Besuner_specc5_2025}. These illustrate the importance of (1) constructing a large sample of LAEs and (2) studying the physical properties of LAEs statistically.

The large spectroscopic dataset collected by the Dark Energy Spectroscopic Instrument (DESI) \citep{Snowmass2013.Levi} offers an opportunity for the compilation of such a statistical sample of LAEs. While the main extragalactic targets of DESI are galaxies at $z<1.6$ and quasars at $0.5 \lesssim z \lesssim 3.5$ \citep{Snowmass2013.Levi, DESI2016a.Science, ELG.TS.Raichoor.2023,QSO.TS.Chaussidon.2023}, some $z>2$ LAEs fall into the color selection of the DESI emission line galaxy sample \citep[ELGs,][]{ELG.TS.Raichoor.2023} as reported in \citet{Lan_2023}. Given the total number of ELGs observed by DESI ($>10^{7}$), even only a sub-percentage of targeted ELGs are LAEs, the final dataset will include at least tens of thousands of LAEs with DESI spectra, a sample sufficient for many scientific explorations, through stacking analysis \citep[e.g.,][]{Davis_2023_50kLAE} and line profile investigations \citep[e.g.,][]{Lan_OII_2024,Mukherjee2025,Uzsoy_env}. The main challenge is to identify LAEs from DESI spectra, given that the current DESI pipeline, Redrock \citep{Redrock.Bailey.2024}, only includes galaxy templates to $z<1.7$. Therefore, those LAEs remain hidden in the DESI dataset and new methods are required to detect them \citep[e.g.,][]{Uzsoy2025}.

In this work, we design a pipeline based on machine learning convolutional neural network (CNN) architecture \citep[e.g.,][]{Lecun_CNN_1990, CNN_LeNet_1998, CNN_Alex_2012} to uncover those hidden LAEs. We construct a visually-identified LAE catalog compiled from DESI data to train the neural network and apply this pipeline to approximately two million DESI spectra in the DR1 dataset \citep{DESI_2025_dr1}. More specifically, we focus on LAEs primarily powered by star-formation with narrow asymmetric Ly$\alpha$ emission lines with weak or no emission from other spectral lines. They tend to have low stellar mass, low metallicity and similar star-formation rate (SFR) compared to the properties of typical star-forming galaxies at similar redshifts \citep[e.g.,][]{Pucha2022}. This population of LAEs dominates the bright end of the Lyman alpha luminosity function and the majority of their redshifts are misidentified by the DESI pipeline. We identify 19,685 LAEs and highlight potential use-cases for scientific explorations with this sample. In addition to the catalog, the developed pipeline can be used for identifying LAEs observed in the DESI-II survey. 

This paper is organized as follows: Section 2 describes the data and visual inspection procedure for constructing the truth table. Section 3 presents the CNN architecture, training procedure and the performance and Section 4 presents the LAE catalog and the basic properties. Finally, Section 5 summarizes the results. We adopt a flat $\Lambda$CDM cosmology with $h=0.6766$ and $\Omega_m=0.30966$ \citep[Planck18;][]{Planck_cosmology_2020}. Galactic extinction corrections are applied to the spectra and magnitudes using \citet{SFD98} dust map and \citet{Fitzpatrick} extinction law. Magnitudes are in AB system.

\begin{figure*}[ht!]
    \centering
    \includegraphics[width=1\linewidth]{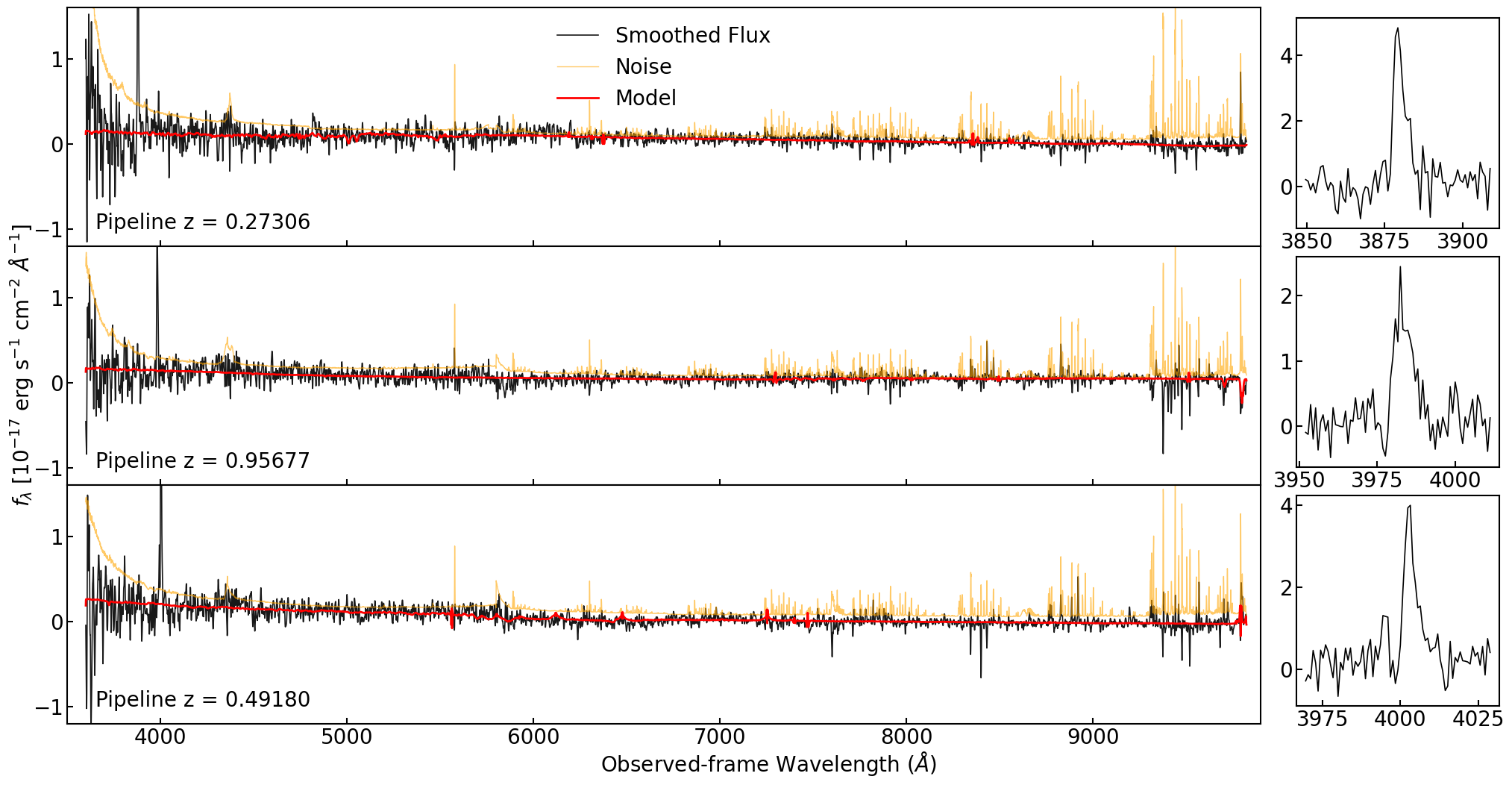}
    \caption{Example spectra of LAEs with Ly$\mathrm{\alpha}$ lines not detected by the DESI pipeline. The black, orange and red lines in each panel are observed spectrum smoothed with a Gaussian filter, the uncertainty, and the DESI best-fit model respectively. The best-fit redshift based on the DESI pipeline is listed at the lower-left corner of each panel. The panels on the right show the zoom-in observed spectra focusing on the Ly$\mathrm{\alpha}$ emission lines. Since Ly$\mathrm{\alpha}$ feature is not captured by the DESI pipeline, the pipeline best-fit redshift is incorrect. The \texttt{SPECTYPE} of these three spectra is either \texttt{QSO} or \texttt{GALAXY}.}
    \label{fig:no_detect_ex}
\end{figure*}

\section{Data}
Constructing a supervised machine learning model for identifying LAEs requires a training dataset to train the model, a validation dataset to optimize hyper-parameters, and a test dataset to quantify the performance of the model \citep[See][for a review]{bishop_pattern_2006, goodfellow_deep_2016}. Therefore, we first need to construct a sample with a large number of LAE spectra. In this section, we present the DESI dataset and the procedure for constructing visually-identified LAE catalog and the dataset to search for LAEs.

\subsection{DESI ELG Spectra}
The Dark Energy Spectroscopic Instrument (DESI) is a large-scale spectroscopic survey designed to obtain redshifts for approximately 40 million galaxies and quasars through spectroscopic observations \citep{Snowmass2013.Levi, DESI2016a.Science}. DESI is mounted on the Mayall 4-meter telescope at the Kitt Peak National Observatory. Its focal plane consists of 5,000 robotic fiber positioners \citep{FocalPlane.Silber.2023}. These fibers are routed to ten spectrographs \citep{Poppett_fiber_2024}, which provide spectral coverage spanning 3600–9800 $\rm \AA$ \citep{DESI2016b.Instr, DESI2022.KP1.Instr, Corrector.Miller.2023}. To enable efficient observations of tens of millions of celestial objects, DESI employs not only an optimized survey operation strategy \citep{SurveyOps.Schlafly.2023}, but also a dedicated target selection \citep{myers_target_2023} and fiber assignment algorithms \citep{FBA.Raichoor.2024}.  In order to process the massive volume of observational data, automated data-reduction pipelines for calibrating raw data to observed flux \citep{Spectro.Pipeline.Guy.2023} and for the determination of source types and redshifts (Redrock) \citep{Redrock.Bailey.2024} are developed.

\begin{figure*}
    \centering
    \includegraphics[width=1\linewidth]{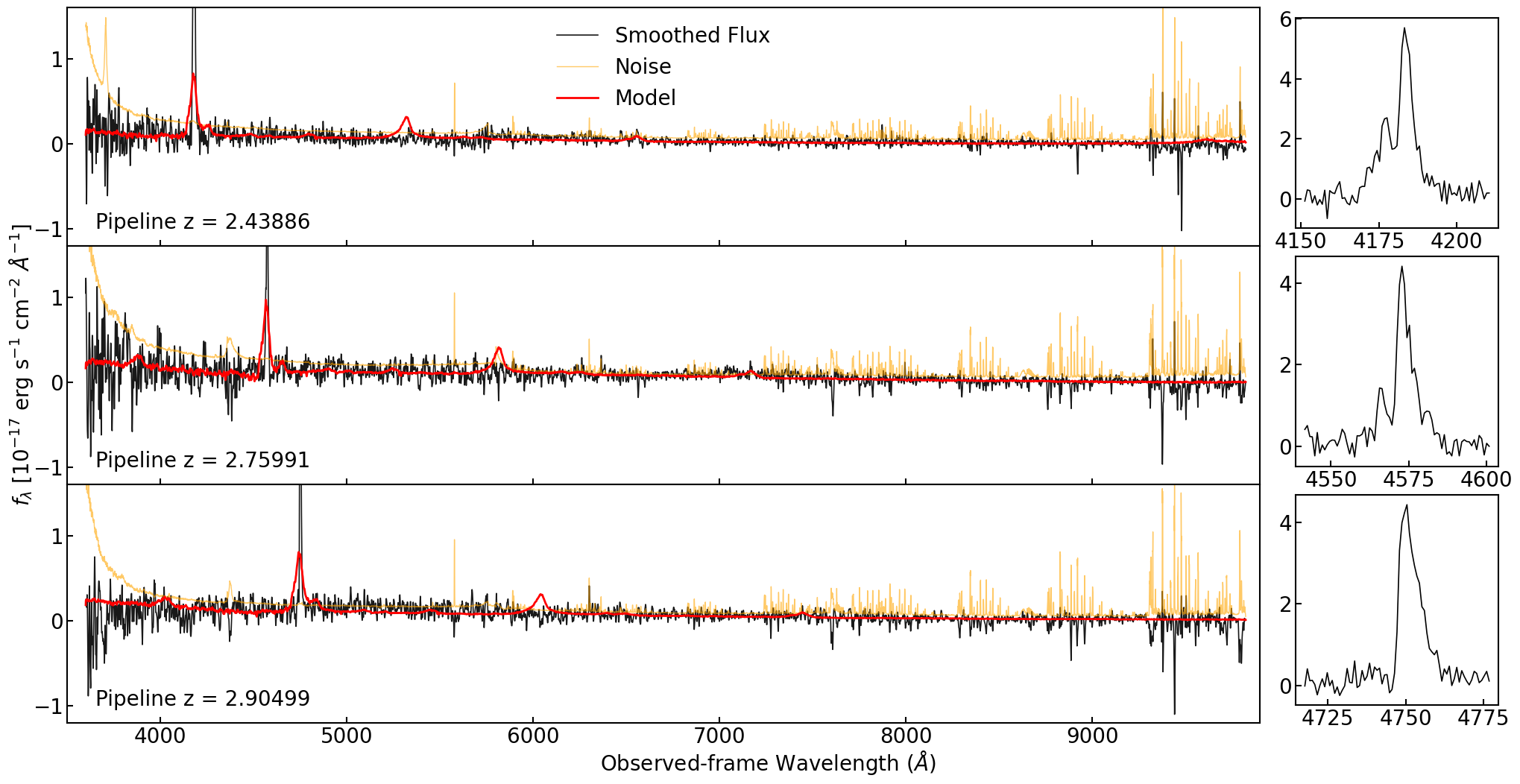}
    \caption{Example spectra of LAEs identified as QSOs by the DESI pipeline. The black, orange and red lines in each panel are observed spectrum smoothed with a Gaussian filter, the uncertainty, and the DESI best-fit model respectively. The best-fit redshift based on the DESI pipeline is listed at the lower-left corner of each panel. The panels on the right show the zoom-in observed spectra focusing on the Ly$\mathrm{\alpha}$ emission lines.
    Although each DESI best-fit model predicts a spurious C~\textsc{iv} emission line, the Ly$\mathrm{\alpha}$ line was identified, yielding a correct redshift. The \texttt{SPECTYPE} of these three spectra is \texttt{QSO}.}
    \label{fig:QSO_ex}
\end{figure*}

\begin{figure*}
    \centering
    \includegraphics[width=1\linewidth]{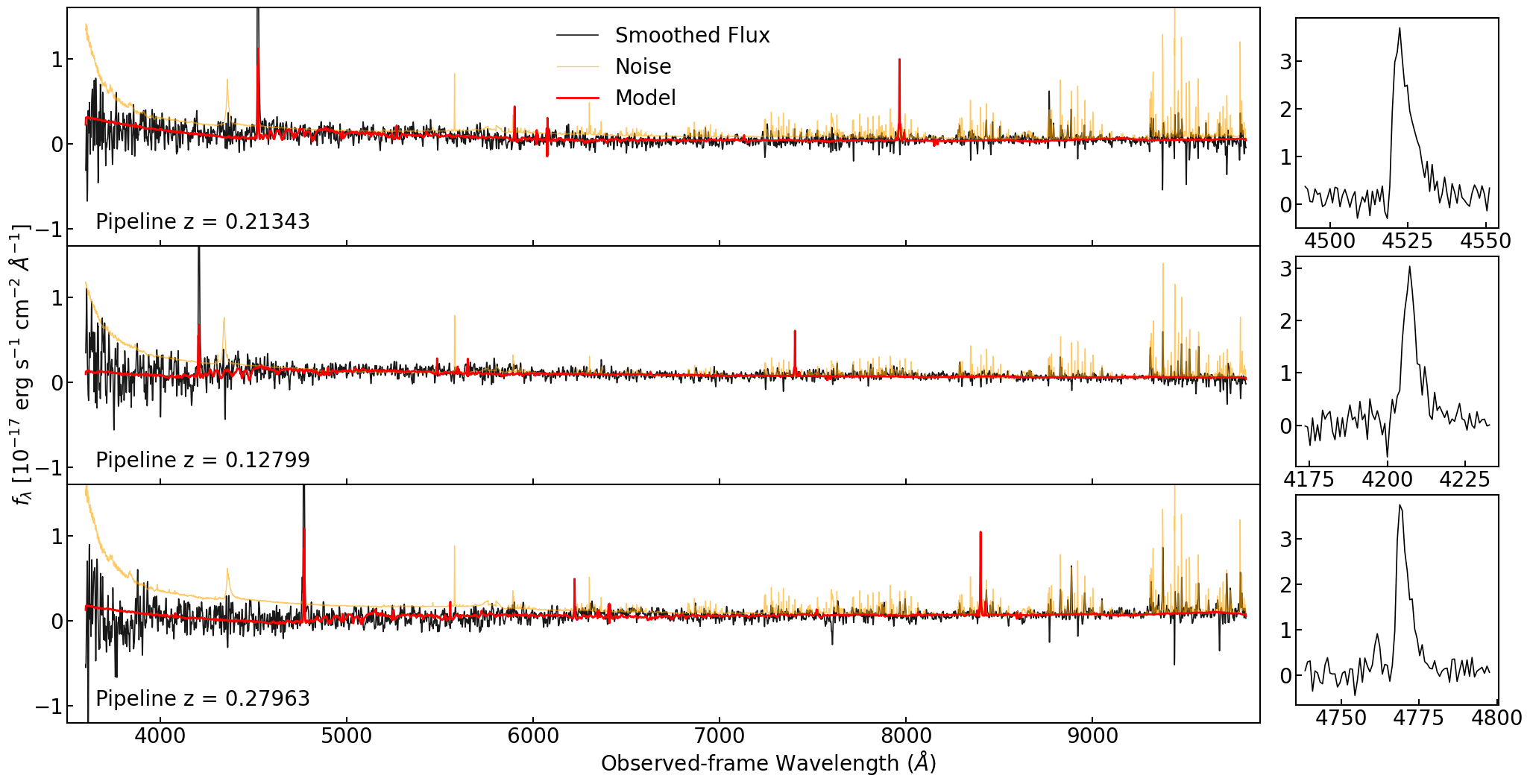}
    \caption{Example spectra of LAEs identified as low-z ELGs by the DESI pipeline. The black, orange and red lines in each panel are observed spectrum smoothed with a Gaussian filter, the uncertainty, and the DESI best-fit model respectively. The best-fit redshift based on the DESI pipeline is listed at the lower-left corner of each panel. The panels on the right show the zoom-in observed spectra focusing on the Ly$\mathrm{\alpha}$ emission lines. The DESI best-fit models tend to predict multiple spurious emission lines, e.g., H$\mathrm{\alpha}$, H$\mathrm{\beta}$, and [O~\textsc{iii}]. Given that the models identify Ly$\mathrm{\alpha}$ emission lines as [O~\textsc{ii}] emission lines, the pipeline redshifts are incorrect (typically, $z_{\text{pipe}}<0.5$). The \texttt{SPECTYPE} of these three spectra is \texttt{GALAXY}.}
    \label{fig:low-z_ex}
\end{figure*}

Among the targets observed by DESI, in this work, we focus on the emission-line galaxies (ELGs) \citep{ELG.TS.Raichoor.2023}, which are selected based on their colors and brightness. The selection is designed to primarily probe star-forming galaxies within the redshift range $0.6 < z < 1.6$ with strong [O~\textsc{ii}] doublet emission lines. Depending on the assigned observational priority, ELG targets are divided into LOw Priority (LOP) and Very-LOw Priority (VLO) samples \citep{ELG.TS.Raichoor.2023}. Both LOP and VLO targets are included in this work.

We use the data collected by the main survey as well as the survey validation phase which was conducted to test the survey hardware, evaluate the performance of the data reduction pipeline, and assess the quality of the final spectra in comparison with expectations \citep{DESI_2024_SV}. This phase included the One-Percent Survey, which observed approximately $1\%$ of the full survey footprint \citep{DESI_2024_SV}. Visual inspections (VI) of the SV spectra \citep{VIQSO.Alexander.2023, Lan_2023} confirmed that the data quality met the pre-defined standards, thereby validating the feasibility of the main survey in delivering precise constraints on cosmological parameters. Moreover, the VI results reveal that some ELG targets are LAEs at redshift $z>2$, motivating this work \citep{Lan_2023}. 
All SV observations were incorporated into the Early Data Release (EDR) \citep{DESI_2024_EDR}. Subsequently, the first year of main survey data, covering the period from May 2021 to June 2022, is included in the First Data Release (DR1) \citep{DESI_2025_dr1}, comprising a total of 18.7 million spectra. Cosmological results with DR1 are reported in \citet[][]{DESI_DR1_cos_2,DESI_DR1_cos_1}. We make use of both EDR and DR1 spectroscopic data for our analysis.

\subsection{Visual Inspection \& Candidate Selection}
\label{subsec:selection}

\textbf{LAE misidentification patterns:} A catalog with true information of LAEs is required in order to train our machine-learning model. To compile such a dataset, we need to first understand the patterns of mis-identifications of LAEs by the DESI pipeline, Redrock. We use the information in the DESI VI catalog for the SV dataset from \citet{Lan_2023} and find 62 VI-confirmed LAEs. They are identified based on the fact that most of the spectra contain only one strong asymmetric emission lines in observed wavelength between 3600$\, \mathrm{\AA}$ and 5000$\, \mathrm{\AA}$. We note that this sample might not be complete because identifying LAEs was not the main task for the SV VI campaign and therefore reporting LAEs was not required. However, we can use these 62 VI-confirmed LAEs to understand how the DESI pipeline mis-identifies their redshifts and source types. We find that there are three misidentification patterns:
\begin{enumerate}
    \item \textbf{No robust redshift detection:} We find that for 23 out of 62 LAEs ($\sim37\%$), the pipeline does not detect Ly$\mathrm{\alpha}$ lines, and therefore no robust redshifts are assigned. Figure~\ref{fig:no_detect_ex} shows example spectra with the black and red lines indicating the DESI observed spectra of LAEs and the Redrock best-fit models respectively. As can be seen, the best-fit models fail to capture the line profiles with model flux being nearly flat across the entire observed spectral regions. The uncertainty arrays are shown by orange lines.
    
    \item \textbf{Mis-identified as QSOs with correct redshifts:} For 
    21 out of 62 LAEs ($\sim34\%$), the pipeline identifies Ly$\mathrm{\alpha}$ lines with correct redshifts. However, the best-fit models show broader Ly$\mathrm{\alpha}$ line profiles with lower amplitude than the observed Ly$\mathrm{\alpha}$ profiles (See Figure~\ref{fig:QSO_ex}). In this case, Redrock determines the LAEs as QSOs (\texttt{SPECTYPE}). This is also reflected in the broad metal emission lines, such as C~\textsc{iv} and C~\textsc{iii}] shown in the best-fit models but not observed in the LAE spectra. This behavior is mainly due to the fact that except for QSO templates, the galaxy templates used by Redrock do not extend to redshift $\sim2$ \citep{Redrock.Bailey.2024}. 
    Therefore, there is no templates to capture narrow Ly$\alpha$ lines of LAEs at such redshifts.
    
    \item \textbf{Mis-identified as [O II] lines and other lines with incorrect redshifts:} 
    For another $\sim21\%$ of LAEs (13/62), the pipeline misidentifies Ly$\mathrm{\alpha}$ lines as [O~\textsc{ii}] doublet lines and yields incorrect redshifts. Most LAEs in this case have assigned redshifts $z=0.1-0.5$ by the pipeline. The best-fit model also predicts other emission lines, e.g., H$_{\alpha}$, [O~\textsc{iii}], and H$_{\beta}$ which are not observed in the spectra as shown in Figure~\ref{fig:low-z_ex}. Among the remaining five LAEs, two Ly$\mathrm{\alpha}$ lines were misidentified as C~\textsc{iii}], another two as C~\textsc{iv}, and the last one as [O~\textsc{iii}].
\end{enumerate}

Based on these mis-identification patterns, we develop two methods to select LAE candidates for follow-up visual inspections. We combine information of line measurements, including flux and noise, from the DESI FastSpecFit catalog presented in \citet{2023ascl.soft08005M} and use both the DESI early data release (EDR) (v3.2 VAC), only focusing on SV sources, and the DESI first data release (DR1) (v2.1 VAC), only with main survey sources. 
\begin{itemize}
    \item \textbf{Method I}: The first method is based on the mis-identification of LAEs as QSOs. In this case, LAEs have correct redshifts from Redrock with no or weak associated metal lines unlike QSOs. We begin by constraining the redshift range of sources with $1.96<z<3.57$ to ensure that the Ly$\mathrm{\alpha}$ line falls within the wavelength coverage of the blue arm of the spectrograph, motivated by the fact that all of the 62 VI-confirmed LAEs found in \citet{Lan_2023} fall in the wavelength coverage of the blue arm. In total, there are 22,204 SV ELGs and 300,796 main-survey ELGs within this redshift range. Then, the potential LAEs were selected with the signal-to-noise (S/N) of the Ly$\mathrm{\alpha}$ line (S/N $> 5$), the S/N of the C~\textsc{iv} line (S/N $< 3$) and the number of pixels of the Ly$\mathrm{\alpha}$ line (\texttt{NPIX $< 50$}) for removing quasars with broad Ly$\alpha$ lines. This method results in 354 SV LAE candidates and 2,279 potential main-survey LAEs.
    
    \item \textbf{Method II}: The second method is based on the mis-identification of Lyman-$\alpha$ lines as [O~\textsc{ii}] lines. 
    First we select sources with Redrock redshift range $0<z<0.49$ to ensure that the [O~\textsc{ii}] line falls within the wavelength coverage of the blue arm. This selection yields 69,190 SV ELGs and 592,872 main-survey ELGs. We identify potential LAE candidates with the S/N of the [O~\textsc{ii}] line (S/N $> 5$), the S/N of the [O~\textsc{iii}] line (S/N $< 5$) and the S/N of the H$\alpha$ line (S/N $< 5$). This method results in 580 selected SV candidates and 3,837 selected main-survey candidates.
\end{itemize}
We note that LAEs without robust redshift measurements are difficult to be extracted based on the available DESI information. However, we will show in Section \ref{sub:performance} that our model can identify such systems.

\begin{table*}
    \centering
    \begin{tabular}{cccccc}
        \hline
        \hline
        Dataset    & Candidate Selection          & QSO (NLAE)             & ELG (NLAE) & Total & Percentage\\
        \hline
        Training   & 2,296 LAEs + 598 NLAEs       & 849                    & 849        & 4,592 & 70$\%$ \\
        \hline
        Validation & 492 LAEs + 128 NLAEs         & 182                    & 182        & 984   & 15$\%$ \\
        \hline
        Test       & 492 LAEs + 128 NLAEs         & 182                    & 182        & 984   & 15$\%$ \\
        \hline
        total      & 3,280 LAEs + 854 NLAEs       & 1,213                  & 1,213      & 6,560 & 100$\%$ \\
        \hline
    \end{tabular}
    \caption{Number of spectra used in training, validation and test datasets. The numbers of sources selected with the methods described in Section~\ref{subsec:selection} are listed in the candidate selection column. Columns QSO (NLAE) and ELG (NLAE) are additional QSOs and ELGs selected using the criteria discussed  in Section~\ref{subsec:tvt_datasets}. The additional QSOs and ELGs are included to better balance the sample and to improve the performance of model to identify features that are not Ly$\alpha$.}
    \label{tab:training-test_set}
\end{table*}

\begin{figure*}
    \centering
    \includegraphics[width=\linewidth]{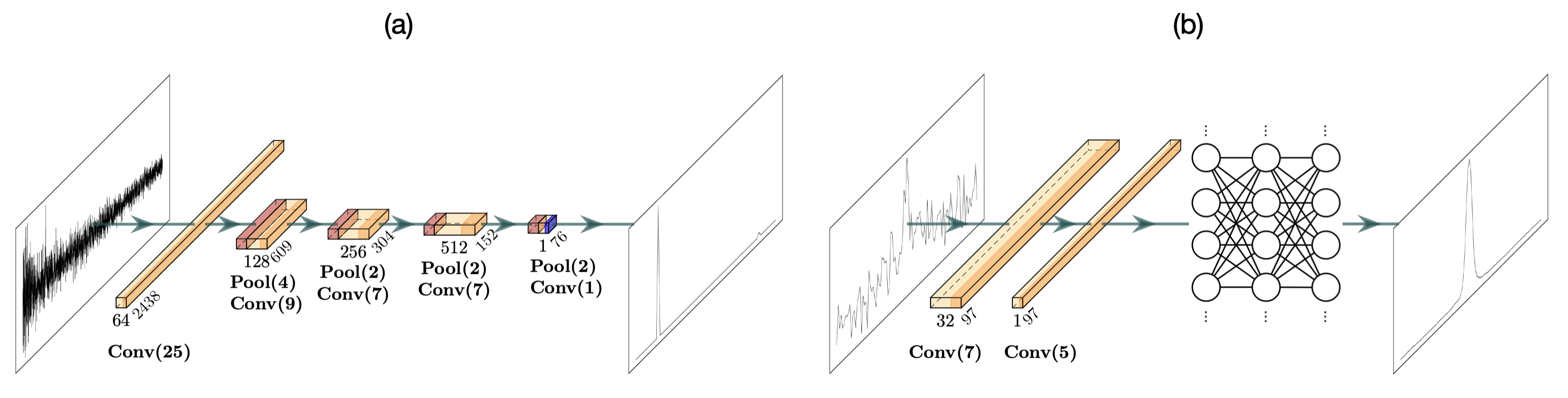}
    \caption{Schematic diagram of the model architecture. (a) Architecture of the first part of our model. Different elements are indicated by different colors ---
    Light orange: convolutional layer (CL), orange: ReLU, dark red: max pooling, and dark magenta: sigmoid. This part has five CLs. Except for the last layer followed by a sigmoid function, each CL is followed by a ReLU function and a max pooling. The number of filters in each CLs are 64, 128, 256, 512, and 1. After successive max-pooling layers, the spectrum size is reduced from 2438 → 609 → 304 → 152 → 76. (b) Architecture of the second part of our model. The same color-code as the one in panel (a). This consists of two CLs and three fully-connected layers (FCLs). Similarly, except for the last FCL followed by a sigmoid function, other layers are followed by ReLU functions. Both input size and output size of this part are 97.}
    \label{fig:CNN}
\end{figure*}

\begin{figure*}
    \centering
    \includegraphics[width=\linewidth]{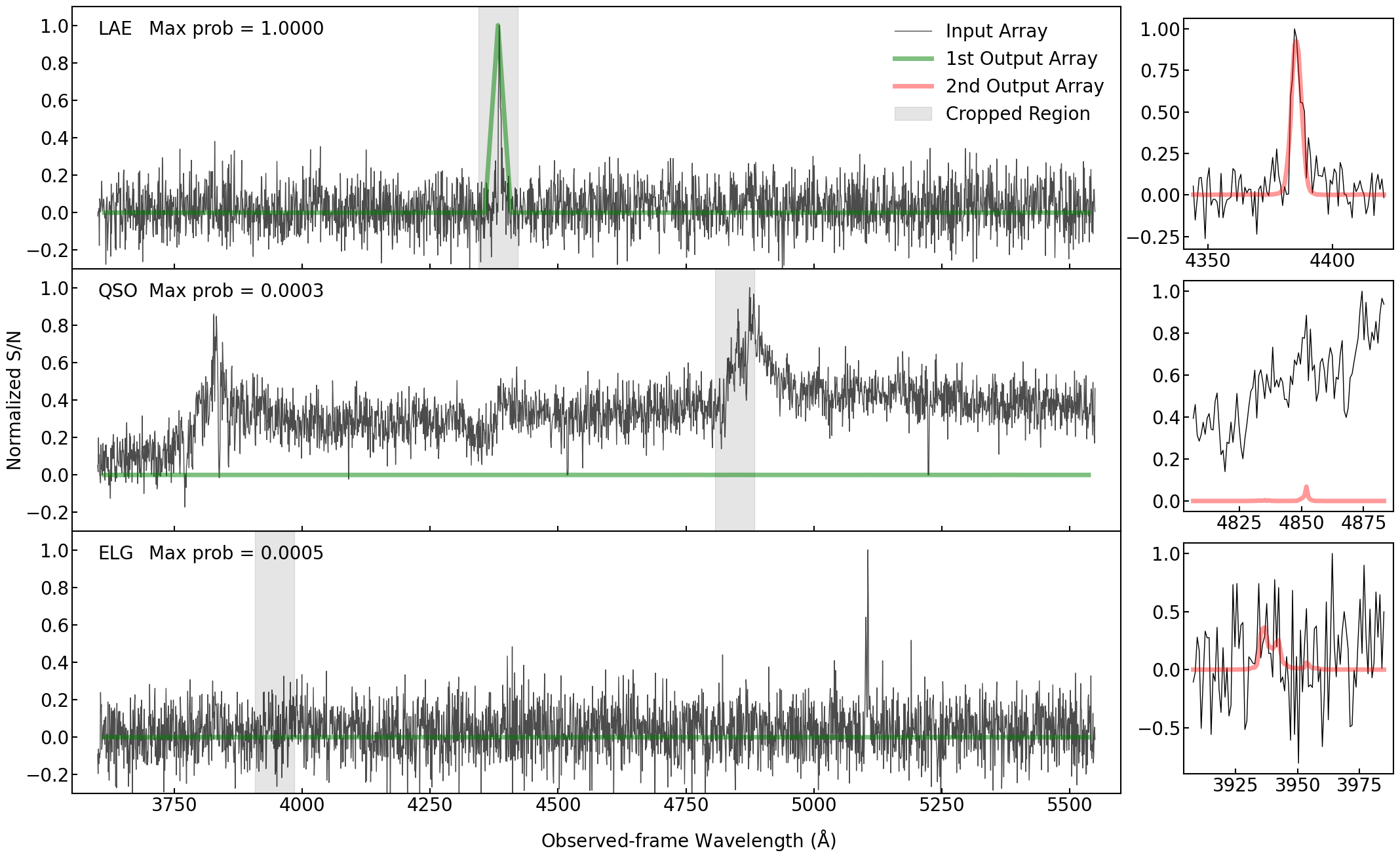}
    \caption{Examples of input data and output arrays from our model. The input data of the model is
    the flux array divided by the error array (S/N array) from $3600$ to $5550\mathrm{\AA}$. We note that for visualization purpose, S/N array (the black curve) is normalized by its maximum value as shown in the figure, while the model uses the original S/N array. The output arrays of the first part of our model indicate the probability of having a Ly$\alpha$ line in each pixel (green curves), and the output arrays of the second part of the model are the Gaussian fitting of Ly$\alpha$ lines in the zoom-in spectral regions identified by the first part (red curves in right panels). Shaded regions in the left panels are the zoom-in segments passed to the second part of the model (right panels). The object type and maximum probability are also shown in the upper-left corner of the main panels.}
    \label{fig:input_output}
\end{figure*}

\textbf{Visual Inspection:} Among the selected LAE candidates based on Method I and II, we visually inspect\footnote{The estimated VI time is $\sim$70 hours.} all the 934 SV LAE candidates and 3,200 main-survey LAE candidates (1,640 via Method I and 1,560 via Method II) to confirm their nature. We consider several spectral features and available DESI information to identify true LAEs, including the presence of spectral lines other than Ly$\alpha$, the shape/width of emission lines, and $\Delta \chi^2$ from Redrock. For each inspected spectrum, we label it with a confidence level from 0 to 2. The higher the level, the more likely the source is a LAE. Basically, the confidence level for each inspected spectrum is primarily based on whether spectral features other than Ly$\alpha$ present in it or not and their apparent strengths. To be more specific, we only assign sources without any other lines significantly observed in the spectra with confidence levels being 2. For spectra with weak observed C~\textsc{iv} or C~\textsc{iii}] lines, we assign the confidence levels being one. Sources with broader Ly$\alpha$ ($\gg25$ pixels in observed frame) and strong spectral features, e.g., C~\textsc{iv}, C~\textsc{iii}], H$\mathrm{\alpha}$, and [O~\textsc{iii}] are labeled as confidence levels being 0. Example spectra with different confidence levels are shown in the Appendix. In this work, since spectra of LAEs powered by star-formation intrinsically have weak C~\textsc{iv} and C~\textsc{iii}] signals \citep[e.g.,][]{Steidel2018, Davis_2023_50kLAE}, we consider sources with confidence levels $\geqslant 1$ to be LAEs. In the end, among 934 SV LAE candidates, there are 756 VI-confirmed LAEs. Similarly, among 3,200 inspected main-survey LAE candidates, there are 2,524 VI-confirmed LAEs.

\subsection{Training, Validation, and Test Datasets}
\label{subsec:tvt_datasets}
When training a model, if the dataset is too imbalanced, it can lead to a biased learning behavior with the model overly inclining to predict the majority class (LAE) while failing to properly identify the minority class (non-LAEs, NLAE) \citep[e.g.,][]{imbalance_class}. In our VI dataset, the number of LAEs (3,280) significantly exceeds that of NLAEs (854), which could hinder the model ability to reliably "ignore" NLAEs. To balance the number of LAEs and NLAEs, we add supplementary spectra in our dataset, consisting of 1,213 quasars and 1,213 low-redshift ELGs. The additional quasar sample is selected using the criteria: (1) \texttt{ZWARN} = 0, (2) $\Delta\chi^2 > 20$, (3) $0.29<z<3.57$, and (4)S/N of any following lines is higher than 5: C~\textsc{iv}, C~\textsc{iii}], or Mg~\textsc{ii}. Similarly, the low-z ELG sample is selected based on the criteria: (1) \texttt{ZWARN} = 0, (2) $\Delta\chi^2 > 20$, (3) $0<z<0.49$, and (4) S/N of H$\mathrm{\alpha}$ and/or [O~\textsc{iii}] lines are higher than 5. 
These two samples are randomly selected from DESI DR1 (v2.1 VAC) dataset.
Finally, our base dataset consists of 6,560 spectra, with a balanced ratio of 1:1 between LAEs and NLAEs.

With this sample, we set the training-validation-test datasets with split ratios $0.70:0.15:0.15$. To ensure that each dataset contains a representative distribution of the samples, we adopt a stratified splitting strategy. The base dataset is divided into six groups based on the data release of the sources, i.e., SV, main survey, or supplementary spectra, and the classes, i.e., LAE or NLAE. Then, for the spectra in each group, we randomly shuffle them and split them according to the ratio above. After applying this stratified splitting procedure, the training, validation, and test datasets contain 4,592, 984, and 984 spectra, respectively. Table \ref{tab:training-test_set} summarizes their composition by source and class. This stratification ensures that the model is exposed to diverse spectra during training while maintaining consistency across validation and test evaluations.

\textbf{The input data format for the model:}
According to the SV LAE analysis, the majority of DESI-detected LAEs have their observed Ly$\alpha$ emission lines within the blue arm of the spectrograph, corresponding to a redshift range of $1.96 < z < 3.57$. 
Therefore, in this work, we focus on LAEs with this redshift range and only use DESI spectra with wavelength range from 3600 to 5550 $\mathrm{\AA}$, covered by the blue arm \citep{DESI2016b.Instr, DESI2022.KP1.Instr}. In addition, to incorporate the spectral noise, instead of using the observed flux directly, we normalize the flux by the corresponding noise and use the dimensionless signal-to-noise ratio (S/N) spectra as the model input.

\section{Detection algorithm}
\subsection{Architecture}
\label{subsec:architecture}
The detection algorithm implemented in this work is based on convolutional neural network (CNN) \citep{Lecun_CNN_1990, CNN_LeNet_1998, CNN_Alex_2012}, an established technique in the field of object detection \citep{Joseph_YOLO_2015, Shaoqing_RCNN_2015}. This technique has also been applied to detect absorption lines in the spectra \citep[e.g.][]{parks_2018_ML, Guo_2019_ML}.
Given the similar nature of detecting Ly$\mathrm{\alpha}$ emission lines (object) from a large dataset, we expect that the technique is well-suited for our goal. We note that a CNN model is also adopted in a recent study by \citet{Mukae2026} to detect Ly$\mathrm{\alpha}$ emission lines from the HETDEX dataset.

Our model consists of two parts:
\begin{itemize}
    \item \textbf{The first part} consists of fully-convolutional layers without any fully-connected layers. Except for the last layer followed by a Sigmoid activation function for probability output, each of the convolution layers is followed by a ReLU activation function \citep{Nair_Hinton_ReLU_2010, CNN_Alex_2012} and a max pooling \citep{CNN_LeNet_1998}, where the latter is used to downscale the size of the input spectrum. To prevent overfitting and accelerate convergence during training process, we also add batch normalization after each convolutional layer \citep{batch_norm}.
    Panel (a) in Figure~\ref{fig:CNN} demonstrates the schematic diagram of this first part. The first part takes the blue-arm S/N spectrum (2438 pixels, 1 pixel $= 0.8 \mathrm{\AA}$) as input and outputs a lower resolution spectrum (76 pixels, 1 pixel $\approx 25 \mathrm{\AA}$), where each output pixel value is from 0 to 1, indicating the probability of the corresponding region of the original spectrum containing a Ly$\mathrm{\alpha}$ emission line. The left panels of Figure~\ref{fig:input_output} show examples of input spectra (black lines) and output arrays (green curves) indicating the possible detection of Ly$\alpha$ emission lines.

    \item \textbf{The second part} is a combination of a slicer and a fitter. The slicer first identifies the pixel from the first part output with the highest probability greater than 0.5, an adopted threshold for determining Ly$\alpha$ detection. It then extracts the corresponding region from the original spectrum, including that pixel and its adjacent neighbors, resulting in a spectrum approximately 75 Å in length. The spectrum is then passed to the fitter. The fitter consists of two convolution layers followed by three fully-connected layers. Similarly, except for the last fully-connected layer followed by a Sigmoid activation function, the others are followed by the ReLU activate functions (also max pooling for the convolution layers). It fits the red peak of LAE Ly$\mathrm{\alpha}$ emission line in the spectrum with a peak-normalized Gaussian. Therefore, the output of the last part is an 97-pixel array containing a Gaussian profile, which allows us to estimate the redshift and the line width of LAEs through finding the peak of Ly$\mathrm{\alpha}$ lines and computing the full-width-half-max (FWHM) of the Gaussian. The right panels of Figure~\ref{fig:input_output} show the input spectra (black lines) of the second part of the model and the output Gaussian profiles (red lines). Since Ly$\mathrm{\alpha}$ is a resonant line, its profile is sensitive to neutral hydrogen density, velocity, and morphology \citep[e.g.,][]{ahn_2004_Lya, Verhamme_2006_Lya, Erb_galphy_2010, Shibuya_LAE_2014, Trainor_2.7LAE_2015, Lya_fitting_ML_2022, Davis_2023_50kLAE, Lya_simulation_2025}. Therefore, our observed Ly$\mathrm{\alpha}$ emission lines are probably redshifted from their $z_{\mathrm{sys}}$. We will further discuss this redshift offset in Section 4.4.

\end{itemize}

The main motivation to adopt a fully convolutional architecture in the first part is that convolution kernels are expected to not only learn the diverse Ly$\alpha$ profiles effectively across the entire input spectra but also reduce the model complexity and therefore the potential of overfitting when comparing to fully-connected layers. During our architecture development phase, we also explored an architecture with fully-connected layers connected after convolutional layers. The model showed overfitting behavior for detecting Ly$\alpha$ lines. After removing all the fully-connected layers, the overfitting behavior disappeared and the model performance improved significantly. This indicates that the generalization of a fully convolutional architecture for detecting Ly$\alpha$ lines is better than the generalization of an architecture with fully-connected layers included. This result motivates the design of our final architecture. The purpose of the second part is to determine the precise location of the Ly$\alpha$ line over $\sim75$ Å wavelength window. Given that this is expected to be a relatively simple task, the second part is designed to have simple components with two convolutional layers and three fully-connected layer. The role of the fully-connected layers here is to map the feature map (convolution of Ly$\alpha$ profiles) to final peak-normalized Gaussian profile.

\subsection{Training Procedure}
\label{subsec:training_procedure}
The training of our CNN model involves several key steps to ensure robust prediction. As described in Section~\ref{subsec:tvt_datasets}, we divide our visually-identified sample and additional sample into training, validation, and test datasets with a 70:15:15 split, employing a stratified strategy to maintain consistency across validation and evaluations.

The model optimization process focuses on 12 hyper-parameters, which are tuned using a multi-stage approach. Specifically, we utilize the \texttt{TPESampler} from the Optuna package \citep{Akiba_optuna_2019} for efficient parameter space exploration and the \texttt{ASHAScheduler} from the Ray package \citep{liaw_tune_2018} to optimize computational resource allocation through early stopping of under-performing trials. Further technical details regarding the specific hyper-parameter configurations, loss function definitions, and the four-stage training sequence are provided in Appendix B.

\begin{table}
\centering
\caption{Confusion matrix (test dataset)}
\begin{tabular}{|c|c|c|c|}
\cline{3-4}
\multicolumn{2}{c|}{} & \multicolumn{2}{c|}{\textbf{Predicted label}} \\ \cline{3-4}
\multicolumn{2}{c|}{} & \textbf{NLAE} & \textbf{LAE} \\ \hline
\multirow{2}{*}{\textbf{True label}} 
    & \textbf{NLAE} & 468 & 24 \\ \cline{2-4}
    & \textbf{LAE}  & 20  & 472 \\ \hline
\end{tabular}
\label{tab:matrix}
\end{table}

\begin{figure}
    \centering
    \includegraphics[width=\linewidth]{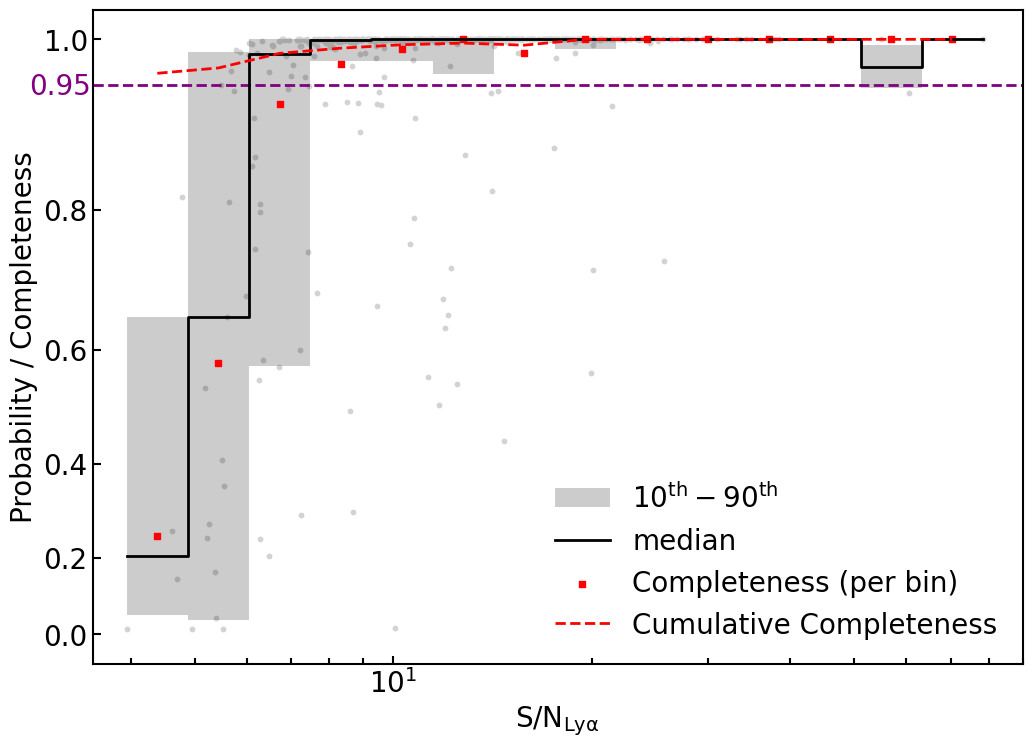}
    \caption{Output probability and detection completeness as a function of Ly$\mathrm{\alpha}$ S/N. Gray dots show predicted probabilities of test LAEs. Red squares and black steps indicate the median probability and completeness, and gray bands show the 10th-90th percentile range of the probabilities. Red dashed line shows the cumulative completeness. Our model detects $>95\%$ of LAEs down to Ly$\mathrm{\alpha}$ S/N =7 . Below this level, detection efficiency drops rapidly, reaching $\sim 60\%$ at Ly$\mathrm{\alpha}$ S/N = 5, and becomes poor at lower Ly$\mathrm{\alpha}$ S/N. Note that the y-axis is in exponential scale to enlarge the upper region.}
    \label{fig:snr_p}
\end{figure}

\begin{figure}
    \centering
    \includegraphics[width=\linewidth]{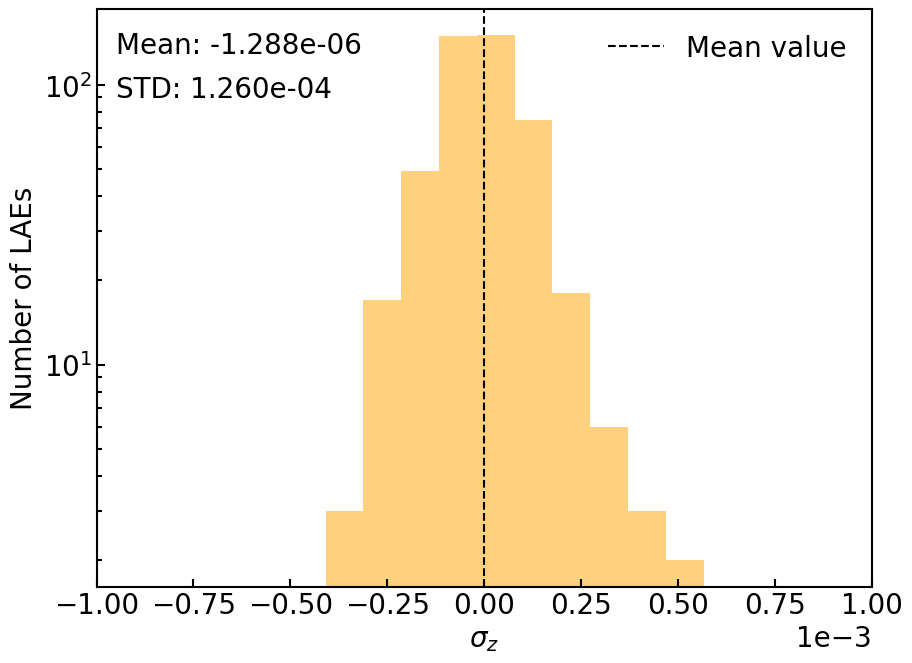}
    \caption{$\sigma_{z}$ distribution of our model.
    The mean value and standard deviation are listed in the upper-left corner with values being $-1.288 \times 10^{-6}$ and $1.260 \times 10^{-4}$ respectively.
    } 
    \label{fig:redshift_test}
\end{figure}

\subsection{Performance}
\label{sub:performance}
To quantify the final performance of our model, we use the confusion matrix in classifying LAEs and NLAEs. Table~\ref{tab:matrix} shows the confusion matrix of the model detection on the test dataset. The diagonal term of the matrix (upper-left and bottom-right) represents correct model prediction (true negative and true positive). The upper-right and bottom-left terms represent incorrect model prediction, false positive (FP) and false negative (FN) respectively. Based on the confusion matrix on the test dataset, the accuracy (TP+TN/Total) is 940/984 = 0.955 [0.941, 0.967], the completeness (TP/TP+FN) is 472/492 = 0.959 [0.938, 0.974], and the purity (TP/TP+FP) is 472/496 = 0.952 [0.929, 0.967]. The intervals shown in the parentheses represent 95\% Wilson confidence intervals \citep{Wilson01061927}.
We also construct the receiver operating characteristic (ROC) curve of our model on the test dataset. 
The area under the curve (AUC) value is 0.988, demonstrating the capability of our model for identifying LAEs.

We further investigate the performance of our model as a function of Ly$\mathrm{\alpha}$ S/N. To estimate Ly$\mathrm{\alpha}$ S/N, we directly sum up the Ly$\alpha$ line flux within $\pm3000 \mathrm{km/s}$ and estimate the corresponding total flux uncertainty by perturbing the spectrum 500 times using its uncertainty array. Figure~\ref{fig:snr_p} shows the relationship between the predicted probabilities from the output of the first part of the model (y-axis) and the Ly$\mathrm{\alpha}$ S/N (x-axis) of the 492 true LAEs in the test dataset. To emphasize the trend, we also bin the data points in 14 equal-size logarithmic intervals and calculate the median value of the probabilities (black lines) and the completeness (red data points) within each bin. 

Overall, the completeness increases with increasing Ly$\mathrm{\alpha}$ S/N. In the range from Ly$\mathrm{\alpha}$ S/N = 4 to Ly$\mathrm{\alpha}$ S/N = 7, the completeness rises from near 0.2 to approximately 0.95. Beyond this range, as the Ly$\mathrm{\alpha}$ S/N continues to increase, the completeness remains nearly $100\%$. In particular, our model can achieve completeness $\gtrsim$ 0.5 at Ly$\mathrm{\alpha}$ S/N $\gtrsim$ 5 and completeness $\gtrsim$ 0.8 at Ly$\mathrm{\alpha}$ S/N $\gtrsim$ 6. We note that the median probabilities drop in the second highest Ly$\mathrm{\alpha}$ S/N bin due to the small sample size. The cumulative completeness (red dashed line) is also shown in Figure~\ref{fig:snr_p}, demonstrating that the cumulative completeness is $>95\%$ over the entire test dataset.

To estimate our model capability of recovering redshifts, we compute redshift error, $\sigma_{z}$, defined as
\begin{equation}
    \sigma_{z}=\frac{z_{\text{model}}-z_{\text{label}}}{1+z_{\text{label}}},
    \label{eq:l/g}
\end{equation}
where $z_{\text{model}}$ refers to the redshift predicted by our model, and $z_{\text{label}}$ refers to the VI-labeled redshifts. Figure~\ref{fig:redshift_test} shows the distribution of $\sigma_{z}$ of the 472 true LAEs with correct prediction in the test dataset. The mean value and standard deviation of $\sigma_{z}$ are listed in the upper-left corner of Figure~\ref{fig:redshift_test} with values being $-1.288\times10^{-6}$ ($-0.4$ km/s) and $1.260\times10^{-4}$ (38 km/s) respectively.

As we only include LAEs in the cases where their Ly$\mathrm{\alpha}$ lines are mis-identified as quasars or  [O~\textsc{ii}] lines in our training data (see Section \ref{subsec:selection} for more details), 
we apply the model to the 23 VI-confirmed SV LAEs without robust redshift measurements (the first pattern in Section \ref{subsec:selection}) to validate whether or not our CNN model can identify those systems. 
We find that only one of them (with Ly$\mathrm{\alpha}$ S/N = 4.61) is not identified by our model. The recovery rate is $22/23 = 95.7\%$, which is consistent with the confusion matrix shown in Table \ref{tab:matrix} ($95.9\%$). This confirms that our model can generally identify LAEs, including systems with a mis-identification pattern which is not included in our training dataset. 

\begin{figure}
    \centering
    \includegraphics[width=\linewidth]{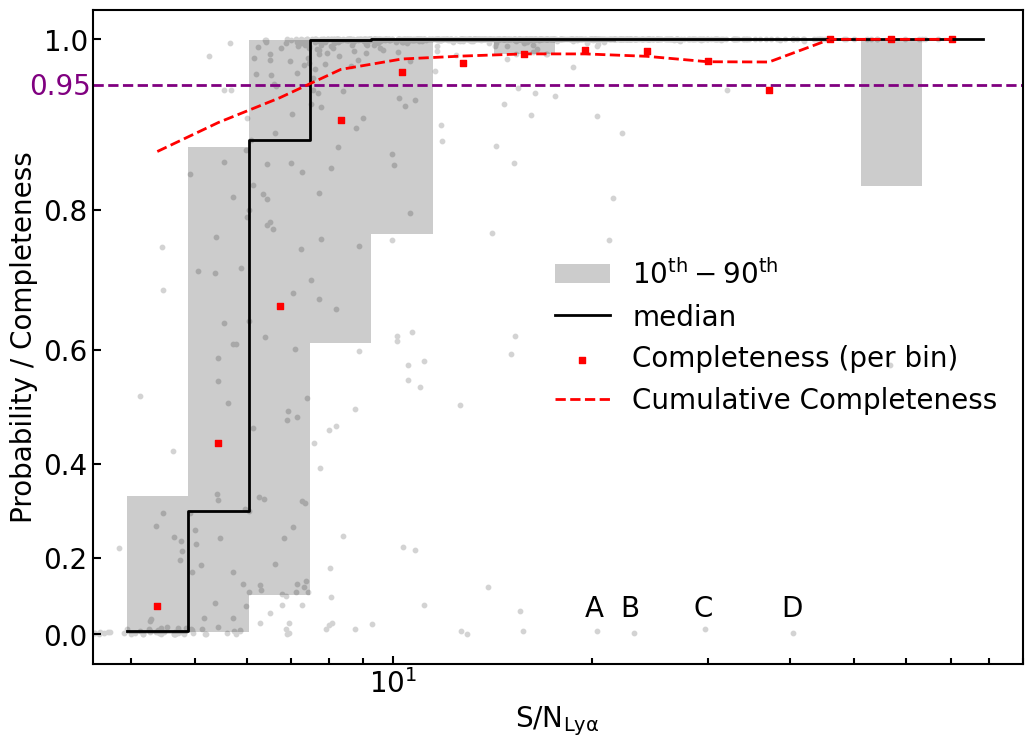}
    \caption{Output probability and detection completeness as a function of Ly$\mathrm{\alpha}$ S/N for DESI-HETDEX LAE sample. Gray dots show predicted probabilities of DESI-HETDEX LAEs. Red squares and black steps indicate the median probability and completeness and grey bands show the 10th-90th percentile range of the probabilities. Red dashed line shows the cumulative completeness. Our model detects $>95\%$ of LAEs down to Ly$\mathrm{\alpha}$ S/N = 8. Below this level, detection efficiency drops rapidly, reaching $\sim 30\%$ at Ly$\mathrm{\alpha}$ S/N = 5. Note that the y-axis is in exponential scale to enlarge the upper region.}
    \label{fig:p_snr_hetdex}
\end{figure}

\subsection{Completeness Estimation Based on DESI-HETDEX LAEs}
\label{sub:DESI-HETDEX}
In addition to estimate the completeness with our test dataset, we estimate the completeness of our model using an independent dataset from the DESI-HETDEX LAE catalog constructed in \citet{DESI-HETDEX_LAE}. This catalog consists of 3,149 targets with follow-up DESI observations of sources observed by the Hobby-Eberly Dark Energy Experiment (HEDTEX), a blind spectroscopic survey for probing LAEs to map large-scale structure at $1.96 < z < 3.57$ \citep{Hill_HETDEX_2008}. The authors visually inspected the DESI spectra and further reported the VI redshifts and the VI qualities in the catalog. The criteria for the highest two confidence levels used for their VI are
\begin{itemize}
   \item[\textbf{4:}] $\geq 2$ secure spectral features and
   \item[\textbf{3:}] one secure spectral feature.
\end{itemize}
While the catalog includes other types of sources, such as stars, AGNs and [O~\textsc{ii}] emitters \citep{Mentuch_2023_HETDEX},
we first use the objects with confidence levels $\geq 3$ and identified as LAEs to test our model completeness for detecting LAEs. The sample consists of 982 HETDEX LAEs. After we apply our model to the DESI spectra of those LAEs, the model can recover $81.3\%$ of them. Similar to Figure~\ref{fig:snr_p}, Figure~\ref{fig:p_snr_hetdex} shows the relationship between the DESI Ly$\mathrm{\alpha}$ $S/N$ and the prediction outcomes with the same symbols and labeling used in Figure~\ref{fig:snr_p}. The overall behavior of the relationship is similar to the one shown in Figure~ \ref{fig:snr_p}, showing that our model is not biased by our training sample and can be applied to an independent dataset directly. However, we note that with the same Ly$\mathrm{\alpha}$ $S/N$, the completeness for DESI-HETDEX sources is lower than the completeness of our test dataset. We also note that the cumulative completeness drop down to $\sim95\%$ at Ly$\mathrm{\alpha}$ $S/N=7$. Additionally, there are four data points with very high DESI Ly$\mathrm{\alpha}$ $S/N$ ($>20$) but low probabilities ($<0.15$), which are labeled A, B, C, and D with their spectra shown in Appendix A. We find that these sources have either broader width of Ly$\mathrm{\alpha}$ emission lines or unusual line profiles with stronger blue peaks (higher ratio of the blue peak amplitude to the red peak amplitude), which are not included in our training dataset. Therefore, our model does not consider them to be LAEs despite the fact the emission line features are significant. In other words, our model not only detects Lyman alpha emission lines but also focuses on narrow LAEs based on the training dataset. Additionally, we also test our model purity by using all the objects with confidence levels $\geq 3$, resulting in 1151 objects (982 LAEs, 131 [O~\textsc{ii}] emitters, 32 AGNs, and 12 other sources). Among 131 [O~\textsc{ii}] emitters, only five of them ($3.8\%$) are misidentified as LAEs by our model. The Ly$\alpha$ lines of 13 AGNs ($40.6\%$) are also identified by our model because they have similar profiles as the profiles of our LAE Ly$\alpha$ lines. Overall, the final purity is $97.7\%$ (798/817).

In principle, we can improve the performance of our model for this DESI-HETDEX sample by including a fraction of them in the training and fine-tune the weights of the hyper-parameters. Nevertheless, given that the goal of this work is to identify LAEs in the DESI sample, we focus on the model trained by the DESI LAE sample. 

\begin{figure*}
    \centering
    \includegraphics[width=\linewidth]{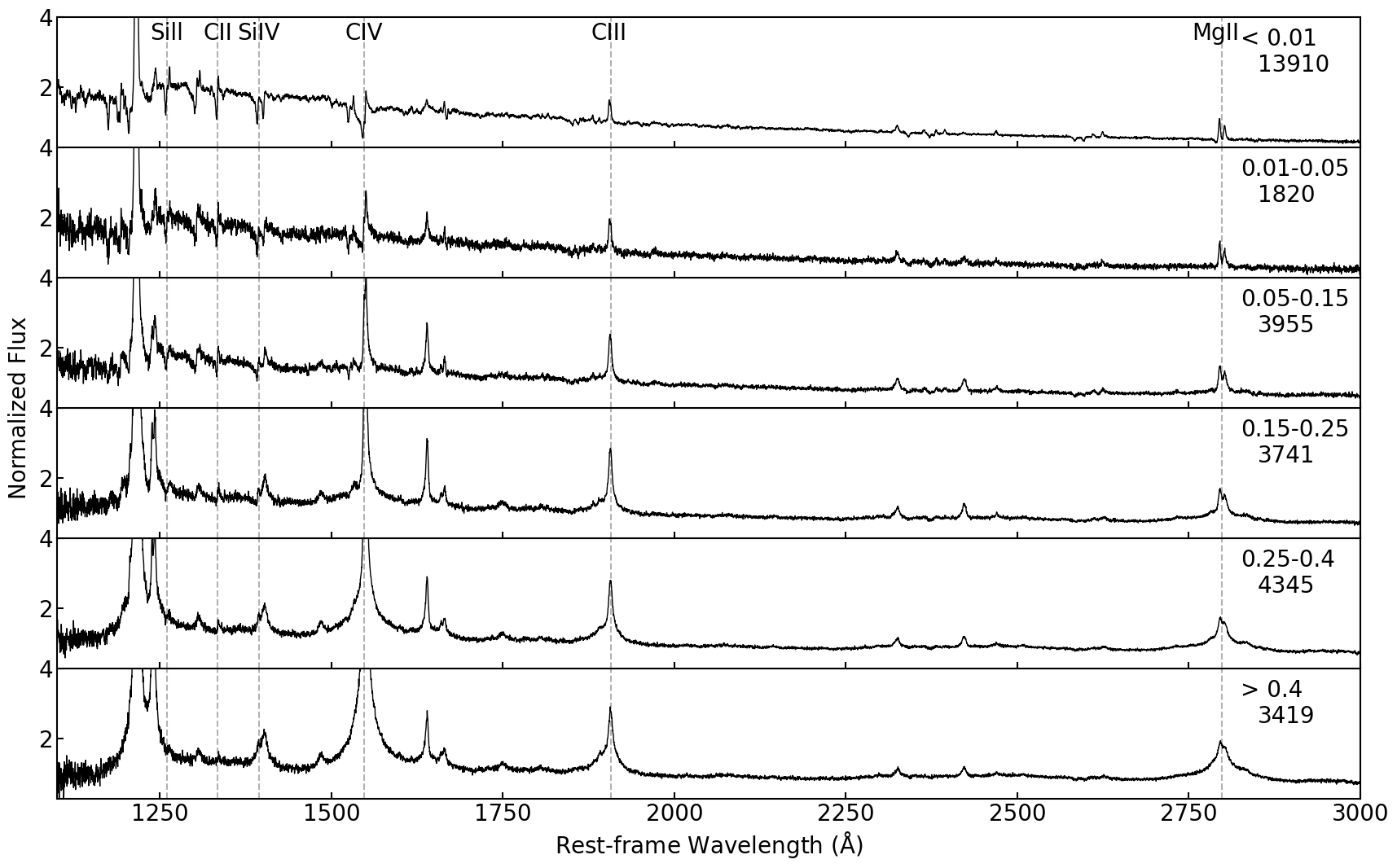}
    \caption{Composite spectra as a function of C~\textsc{iv}/Ly$\alpha$ flux ratio. The range in the upper-right corner of each panel shows the flux ratio bin. Several apparent changes in spectral lines are observed, such as the increasing strength of C~\textsc{iv}, C~\textsc{iii}], and Mg~\textsc{ii}, the disappearing P-Cygni profiles of Si~\textsc{ii} and C~\textsc{iv}, and the transformation of Si~\textsc{iv} from absorption to emission. We set C~\textsc{iv}/Ly$\alpha$ flux ratio = 0.15 as the threshold for separating star-forming LAEs and AGNs.}
    \label{fig:composite_SNR}
\end{figure*}

\section{DESI Lyman Alpha Emitter Catalog}
With this trained model, we now apply it to the DESI galaxy spectra to uncover unidentified LAEs. 
We select sources, targeted as ELGs in the DESI DR1 dataset, using two complementary criteria:
\begin{enumerate}
    \item $\Delta \chi^{2}<20$ or 
    \item $\Delta \chi^{2}\geqslant20$ and o2c $< 0.9$
\end{enumerate}
where $\Delta \chi^{2}$ is defined as the difference in $\chi^{2}$ between the best-fit and the second-best-fit models, serving as a measure of how strongly the best-fit model is favored \citep{Redrock.Bailey.2024} and o2c is defined as $\rm log_{10} [O~\textsc{ii}]\,S/N+0.2\rm \, log_{10}\Delta \chi^{2}$ \citep{Lan_2023, ELG.TS.Raichoor.2023}. The first selection covers the ELG spectra with relatively low confidence of their redshift determination from Redrock using $\Delta \chi^{2}$ as an indicator. The total number of galaxies satisfying the first selection is 1,343,617. The second criterion selects sources with spectra that have relatively high confidence of their redshift determination from Redrock, but do not pass the signal-to-noise criterion of [O~\textsc{ii}] emission lines for ELGs \citep{Lan_2023, ELG.TS.Raichoor.2023}. This selection includes 672,711 galaxies. 
We run our model with a single A100 GPU 
and it only takes $\sim12$ minutes to process $\sim 2$ million spectra.
The process time includes the pre-processing but not obtaining or downloading spectrum files.
From this sample, we identify 33,983 LAE candidates, consisting of 17,251 and 16,732 systems from the first and second selection criteria respectively. 

\subsection{Post-processing for Reducing Contamination and Final Catalog}
\label{sub:final_catalog}
To further validate the purity of the detected LAE candidates, we visually inspect spectra of 340 LAE candidates. The results show that the purity of the sample selected with $\Delta \chi^{2}<20$ is $\sim95\%$, consistent with the result based on our test dataset. On the other hand, $\sim50\%$ of the LAE candidates with $\Delta \chi^{2}\geqslant20$ and $\rm o2c<0.9$ show metal emission lines, indicating that a non-negligible fraction of sources in this sample tends to be AGNs. Via visual inspection, we find that the shape and the width of narrow Ly$\mathrm{\alpha}$ emission lines from AGNs are indistinguishable from those of the LAE candidates having only apparent Ly$\mathrm{\alpha}$ emission lines. In addition, we find that $\sim 3\%$ of the detected LAE candidates with $\Delta \chi^{2}\geqslant20$ and $\rm o2c<0.9$ are also AGNs with their C~\textsc{iv} or He~\textsc{ii} emission lines mis-identified as Ly$\mathrm{\alpha}$ emission lines by our model.

First, in order to remove the sources with incorrect redshift prediction, we assume the identified Ly$\mathrm{\alpha}$ features are O~\textsc{vi}, Ly$\beta$, N~\textsc{v}, C~\textsc{iv}, He~\textsc{ii}, or C~\textsc{iii}] emission lines and compute the corresponding new "correct" redshifts. Then at new redshifts, we compute potential flux and S/N of each of the following spectral features: {Ly$\alpha$, C~\textsc{iv}, He~\textsc{ii}, C~\textsc{iii}, Mg~\textsc{ii}]} (if they fall in DESI wavelength range) by directly summing up spectrum within $\pm1000 \, \mathrm{km/s}$. Therefore, for each spectrum and for each assumed feature, we have $3-4$ individual S/N estimations. Sources with any S/N estimation higher than 3 are considered to have incorrect line identification by our model and are removed from our sample. This removes 2,793 sources (8.2\%).

Then, in order to separate the LAE candidates with Ly$\alpha$ emission powered by star-formation and AGNs, we adopt the following approach. 
First, since the remaining objects pass the first stage selection, we assume that the redshifts predicted by our model are correct. We then compute Ly$\alpha$ flux and potential C~\textsc{iv} flux by directly summing up spectrum within $\pm3000 \, \mathrm{km/s}$ of the two lines. The relatively wide window is to capture possible broader emission lines from AGNs. If one spectrum has higher flux ratio C~\textsc{iv}/Ly$\alpha$ along with broader emission lines, it is considered to be an AGN spectrum. To determine flux ratio threshold, we obtain composite spectra of the LAE candidates as a function of the CIV/Ly$\alpha$ flux ratio. To do so, we first shift every LAE spectrum to the rest frame and resample them to a common grid ranging from $791.58\mathrm{\AA}$ to $3311.32\mathrm{\AA}$ with pixel size $=0.267 \mathrm{\AA}$, determined by the DESI spectral resolution (0.8\AA/(1+z)) at $z=2$. We normalize every spectrum by calculating the median value of the flux at $1420 < \lambda < 1520 \mathrm{\AA}$, $1680 < \lambda < 1850 \mathrm{\AA}$, and $1930 < \lambda < 2300 \mathrm{\AA}$ and dividing the spectra by the median value. Finally, we combine all of the normalized spectra with a median estimator to obtain the composite spectrum. The results are shown in Figure~\ref{fig:composite_SNR}. The flux ratio range and the number of spectra used to produce the composite spectrum are shown on the upper-right corner of each panel. We notice that the Si~\textsc{iv} absorption features transition to emission features around flux ratio = 0.15. Additionally, the broad component of C~\textsc{iv}, C~\textsc{iii}], and Mg~\textsc{ii} emission lines also increase significantly around flux ratio = 0.15. This transition suggests that AGNs begin to dominate the high ratio subsample, replacing the stellar-wind–driven P-Cygni profiles with broad emission line profiles from AGNs. We also note that the composite spectra of the systems with flux ratio$>0.15$ are similar to the spectra of Type II quasars \citep[e.g.,][]{Alexandroff2013}. Therefore, we consider the CIV/Ly$\alpha$ flux ratio = 0.15 a viable empirical separation between LAEs powered by star-formation and by AGNs. In this second stage, we identify 11,505 objects with flux ratio $>$ 0.15.

In the following, we denote the LAE candidates with CIV/Ly$\alpha$ flux ratios $\leq0.15$ as LAEs, and the LAE candidates with ratios $>0.15$ as AGNs.

\begin{figure*}[ht!]
    \centering
    \includegraphics[width=0.75\linewidth]{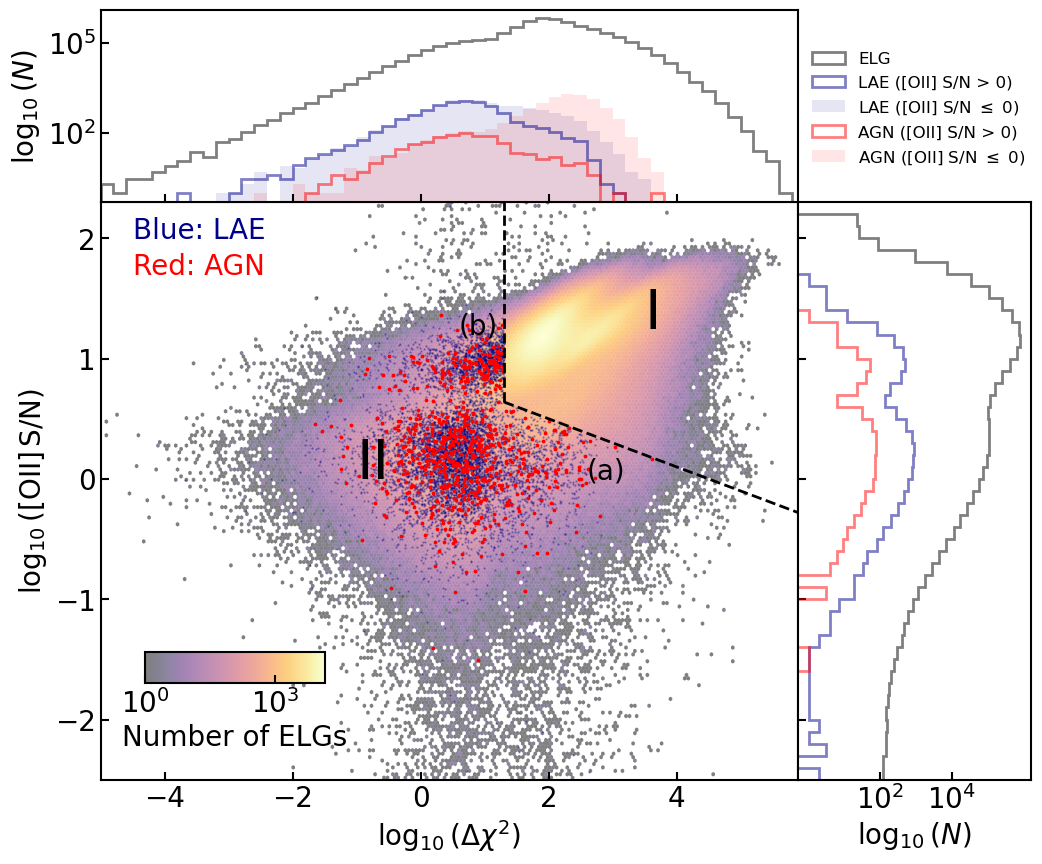}
    \caption{$\log_{10}\Delta\chi^{2}$ vs $\log_{10}\mathrm{[OII]}$ S/N distribution of LAE candidates. Middle panel: Background colormap represents all of the ELGs in DESI DR1 main survey except for the sources with [O~\textsc{ii}] S/N $\leq$ 0. Dark blue and red data points represent the LAE sample (flux ratio $\leq$ 0.15) and the AGN sample (flux ratio $>$ 0.15) in our sample, respectively. Dashed line (a) and (b) indicate $\Delta\chi^{2}=20$ and $\rm log_{10} [O~\textsc{ii}]\,S/N+0.2\rm \, log_{10}\Delta \chi^{2}=0.9$. Region II is the region used to search for LAEs. The right and top panels show the number distributions of the total ELGs, LAEs, and AGNs in black, blue, and red respectively.}
    \label{fig:OII_deltachi2}
\end{figure*}

Figure~\ref{fig:OII_deltachi2} shows the distributions of the LAE sample and the AGN sample in $\log_{10}(\Delta\chi^{2})$ vs $\rm \log_{10}([O~\textsc{ii}]\,S/N)$ space. The blue and red data points represent the LAEs and AGNs, respectively. The background color map shows the distribution of all ELGs in DESI DR1 main survey. 

\textbf{The final catalog:}
Our final catalog consists of two samples: (1) 19,685 LAEs with 14,981 from $\Delta \chi^{2}<20$ selection and 4,704 from $\Delta \chi^{2}\geqslant20$ selection. (2) 11,505 AGNs with 1,453 from $\Delta \chi^{2}<20$ selection and 10,052 from $\Delta \chi^{2}\geqslant20$ selection. Given that our goal is to detect LAEs with Ly$\alpha$ emission lines powered by star-formation, the population that is mostly missed by Redrock, in the following, we focus on the exploration of the LAE sample. We provide same measurements of the AGN sample in Appendix C. However, we note that due to the fact that our training, validation, and testing datasets are constructed for the LAE population, the completeness and purity estimations presented in Section 3.3 and 3.4 do not reflect the completeness and purity of the AGN sample.

To further quantify the purity of the final LAE catalog, we visually inspected 500 random sources from 19,685 LAEs. We identified two systems with incorrect redshifts with CIV lines misidentified as Lya and seven systems with correct redshifts but with apparent CIV, HeII and CIII] emission line features which we considered them to be AGN signatures. However, these seven systems all pass the CIV/Ly$\alpha$ flux ratio cut $<0.15$. If we consider these 9 systems to be incorrect identifications, this leads to the estimated purity being 98.2\% (491/500) with 95\% Wilson CL being [96.6\%, 99.1\%]. If we only consider the two systems with incorrect redshifts to be incorrect identifications, the estimated purity is 99.6\% (498/500) with 95\% Wilson CL being [98.6\%, 99.9\%].

\begin{figure*}
    \centering
    \includegraphics[width=0.9\linewidth]{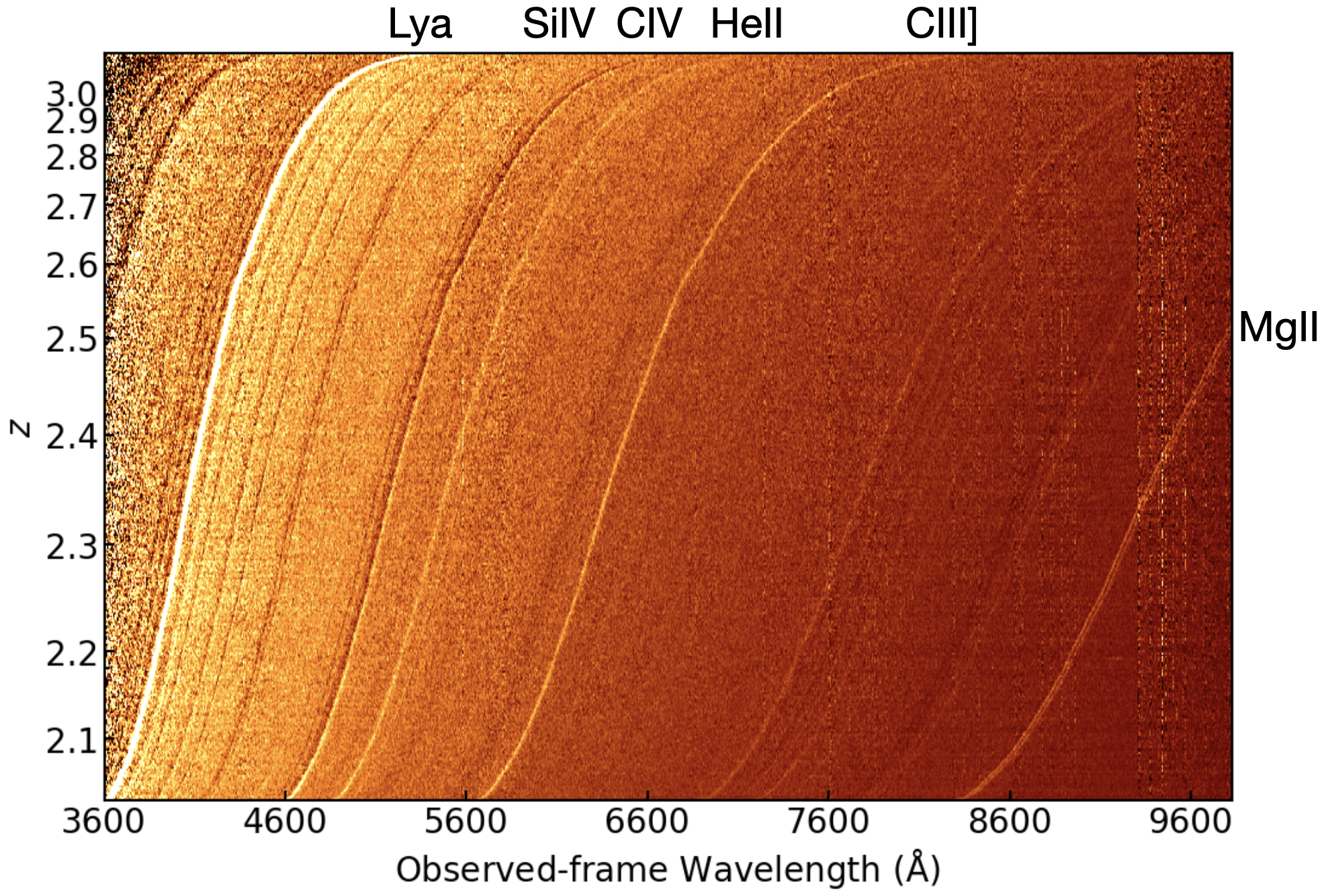}
    \caption{Visualization of LAEs in our catalog in observed frame. The y-axis is the redshift predicted by our model, and the x-axis is the observed frame wavelength. 
    The color code indicates the strength of intensity in each pixel; the brighter the pixel, the higher the intensity. 
    The figure shows the spectra of $5\%$ of the total sources with a 2D Gaussian filter applied to smooth and enhance the signal-to-noise ratio of the map. 
    In addition to the Ly$\alpha$ emission lines, many other metal lines, labeled on the edge, can be seen.}
    \label{fig:LAE_demonstration}
\end{figure*}

Figure~\ref{fig:LAE_demonstration} visualizes the LAE sample with the x-axis being the observed frame wavelength, and the y-axis being redshifts predicted by our CNN model. The colors represent the relative flux of smoothed and normalized LAE spectra. The figure shows only $5\%$ of the total spectra with a 2D Gaussian filter applied to smooth and enhance the signal-to-noise ratio of the map. In the figure, many weak emission and absorption lines, including Mg~\textsc{ii}, C~\textsc{iv}, C~\textsc{iii}], Fe~\textsc{ii}, can be observed, illustrating the rich spectral information in this dataset. Some systematic features, including sky residuals at specific observed wavelengths, can also be observed. 

\begin{figure}
    \centering
    \includegraphics[width=0.9\linewidth]{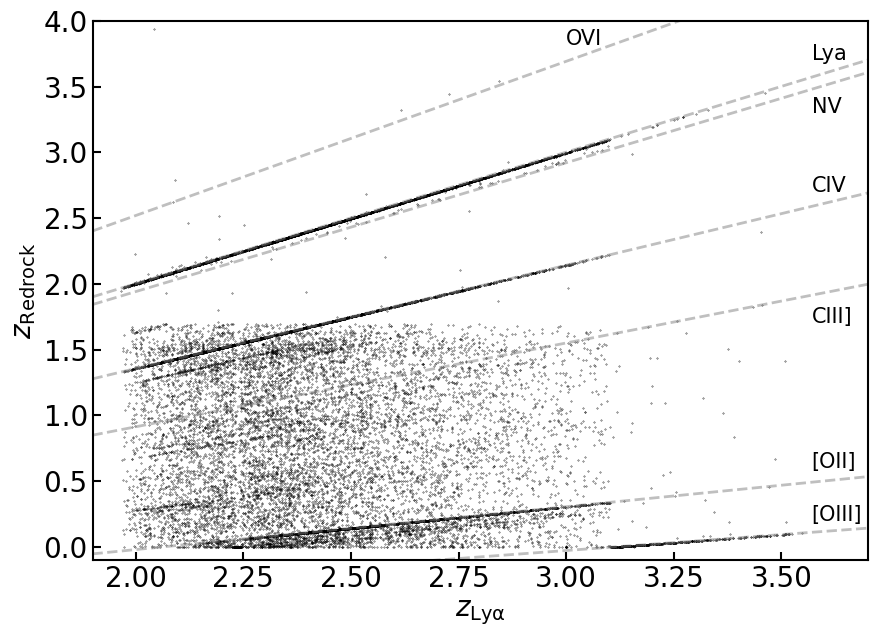}
    \caption{Distribution of LAE redshifts predicted by our model ($z_{\mathrm{Ly\alpha}}$) and Redrock ($z_{\mathrm{Redrock}}$).
    The alignments of data points indicate that a spectral line has been systematically misidentified as another spectral line except the one with unit slope. On the other hand, when the Ly$\alpha$ line is missed by the pipeline, Redrock redshifts distribute randomly within the range allowed by the pipeline. Several straight lines are highlighted using gray dashed lines, and the spectral lines to which Ly$\alpha$ are misidentified are labeled on the right.}
    \label{fig:zm_zr}
\end{figure}

By comparing the redshifts predicted by our model and Redrock, we explore the misidentification patterns by Redrock.  Figure~\ref{fig:zm_zr} shows the LAE redshifts predicted by our model (x-axis) and by Redrock (y-axis). When Ly$\alpha$ line is systematically misidentified as another spectral line by Redrock, the corresponding data points align along a straight line in Figure~\ref{fig:zm_zr}. On the other hand, if no specific spectral line is captured by Redrock, the data points are expected to distributed randomly (within the redshift range allowed by the pipeline). As shown in  Figure~\ref{fig:zm_zr}, one can observe that Ly$\alpha$ lines tend to be misidentified as various emission lines, highlighted on the right of the figure, by Redrock. 
Here we summarize top four systematic misidentification cases among 19,685 LAEs:
\begin{itemize}
    \item $\sim$3,500 with correct redshift but classified as QSO (17.8\%),
    \item $\sim$2,900 misidentified as C~\textsc{iv} (14.7\%),
    \item $\sim$1,700 misidentified as [O~\textsc{ii}] (8.6\%),
    \item $\sim$520 misidentified as [O~\textsc{iii}] (2.6\%).
\end{itemize}
For the remaining $\sim$56\% of the sources, the pipeline does not detect most of the Ly$\alpha$ lines, assigning random redshifts to the sources. 
In other words, $\sim82\%$ of the sources in the LAE sample do not have correct redshifts from Redrock.

With this catalog, in the following, we explore the basic properties of LAEs, including their observed color distributions, redshifts and Ly$\alpha$ luminosity and spectral information in the spectra.

\subsection{LAE Color Distributions}
\begin{figure}[ht!]
    \centering
    \includegraphics[width=\linewidth]{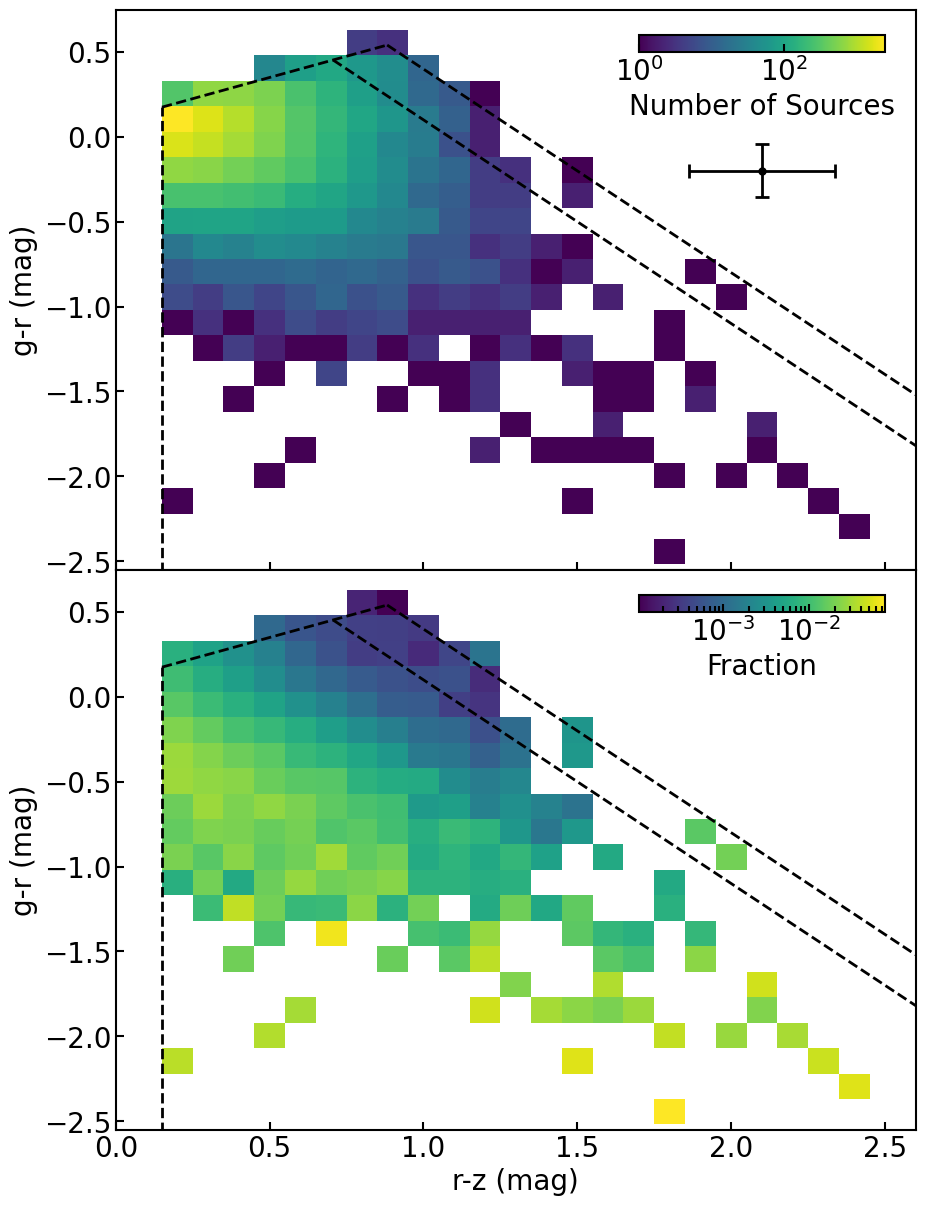}
    \caption{Upper panel: Number of LAEs as a function of g-r and r-z. Lower panel: Ratio of number of LAEs to total number of ELGs in DESI DR1 main survey as a function of g-r and r-z. The typical $1 \sigma$ uncertainties of g-r and r-z are 0.155 and 0.237 mag, respectively, as shown by the error bar.}
    \label{fig:color_fraction}
\end{figure}

We first investigate the color distribution of LAEs in our LAE catalog and compare with DESI ELG target selection.  
Figure~\ref{fig:color_fraction} shows $g-r$ vs $r-z$ distribution with the upper panel illustrating the number of LAEs and the lower panel showing the fraction of LAEs within a given color-color bin. 
The color codes in the two panels correspond, respectively, to the total number of LAEs in each color-color bin and to the ratio of this number to the total number of ELGs (from the main survey, dark program).
For comparison, we also plot the DESI ELG target selection as shown by the dashed lines, including low priority (LOP) and very low priority (VLO) target selections. 
One can observe that for sources with bluer $g-r$ and $r-z$ colors, the fraction of LAEs is larger.

\subsection{Redshift \& Ly$\mathrm{\alpha}$ Luminosity}
\label{Lya_luminosity}
\begin{figure}
    \centering
    \includegraphics[width=\linewidth]{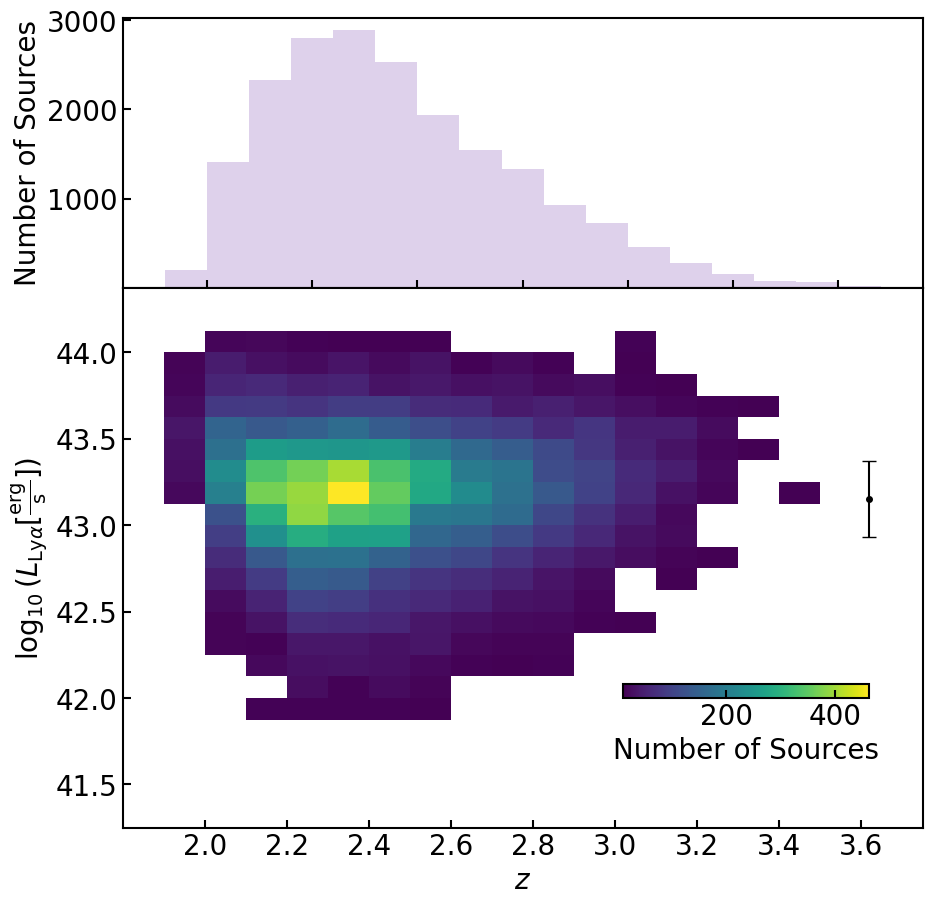}
    \caption{Redshift Distribution and Ly$\mathrm{\alpha}$ luminosity as a function of redshift. 
    The color code in the lower panel indicates the number of LAEs within the bin. The error bar ($1 \sigma$) demonstrates the typical uncertainty (0.220 dex).
    }
    \label{fig:luminosity_redshift}
\end{figure}

We further estimate the Ly$\alpha$ emission line luminosity via directly summing up all the luminosity across the Ly$\alpha$ spectral line region range. 
Figure~\ref{fig:luminosity_redshift} shows the redshift and Ly$\alpha$ luminosity distribution of the LAEs. The upper panel shows the LAE number as a function of redshift and the lower panel shows the luminosity and redshift distribution with colors indicating the number of systems. 
The number of LAEs increases from $z\sim1.9$, reaches the peak at $z\sim2.4$ and declines towards higher redshifts. This behavior is consistent with the combination of observed volume, the sensitivity of the spectra as a function of wavelength and flux limit. First, the DESI spectra exhibit larger uncertainties at the blue end ($3600-4000\mathrm{\AA}$) of the blue arm due to relatively low instrument throughput and atmospheric extinction \citep{DESI2016b.Instr, DESI2022.KP1.Instr, Poppett_fiber_2024}. This reduces the detection efficiency of our model at $z = 1.96 - 2.29$. In addition, with the same sky coverage, the volume at $z\sim2$ is smaller than the volume at $z\sim2.4$. The combination of the two explains why the number of detected LAEs at $z\sim2$ is smaller than the number of detected LAEs at $z\sim2.4$. For higher redshifts, given that the effective exposure time for all the ELG targets is $\sim1,000s$, it limits the detectability of LAEs with lower intrinsic luminosity at higher redshifts. Nevertheless, our catalog covers LAEs at the bright end of Ly$\alpha$ luminosity function \citep{Ouchi_LF_2008, Konno_LF_2016} with $>10^{43}\, \rm erg/s$ and can be used to explore the properties of this population.

\begin{figure*}
    \centering
    \includegraphics[width=\linewidth]{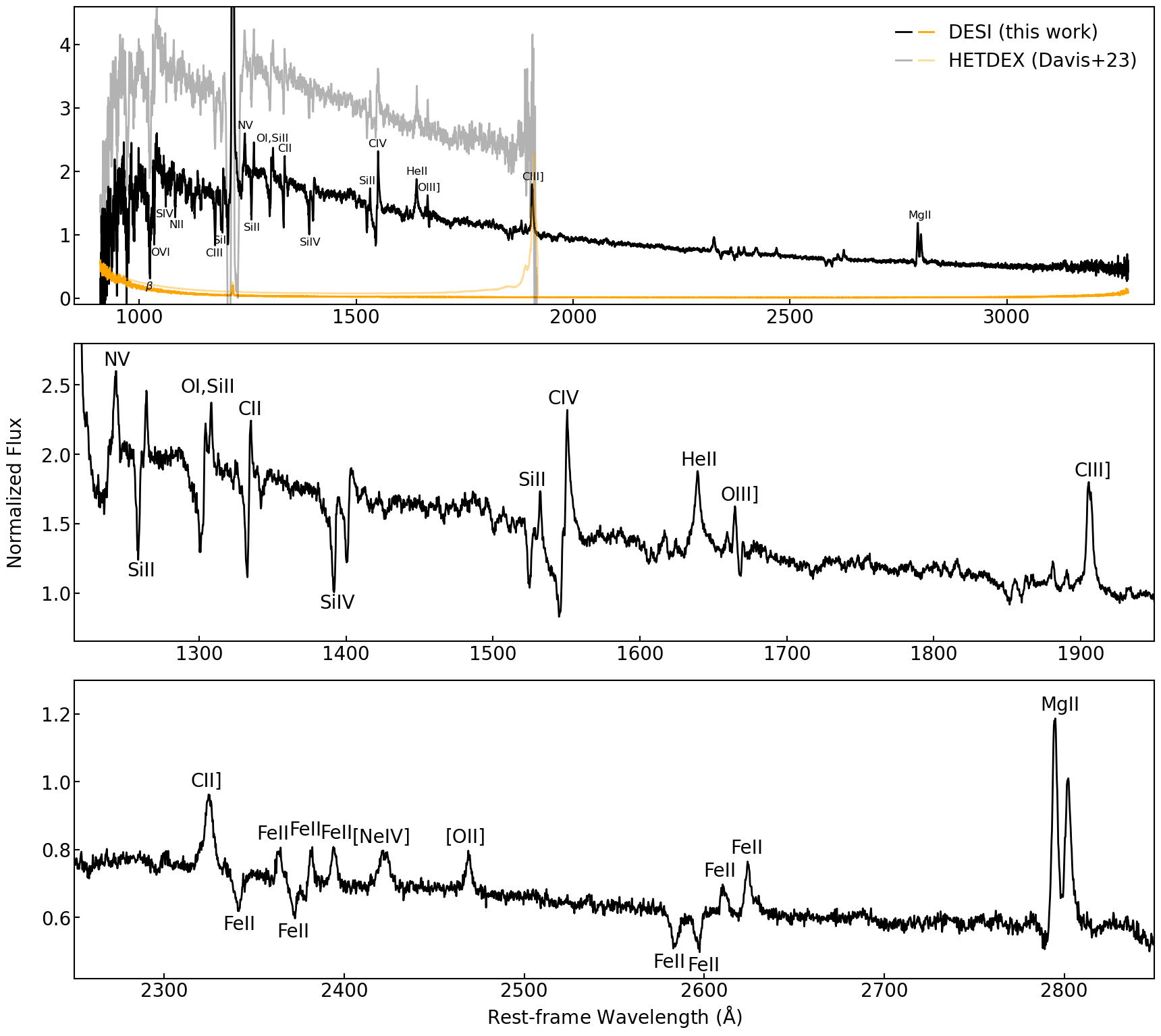}
    \caption{Composite spectrum of 19,685 LAEs. The black and orange curves are the normalized flux of the composite spectrum and the uncertainty estimated from 500 bootstrap realizations, respectively. The gray and faint orange curves in the upper panel are the composite spectrum from \cite{Davis_2023_50kLAE}, which is re-normalized and offset by a constant value of 1.5 for visualization purpose, and its uncertainty. The middle panel is a zoom-in version of the upper panel, focusing on $1215 < \lambda_{\text{rest}} < 1950 \mathrm{\AA}$. The lower panel is a zoom-in version of the upper panel, focusing on $2250 < \lambda_{\text{rest}} < 2850 \mathrm{\AA}$ and illustrating Mg~\textsc{ii} and the weak absorption and emission lines. We note that the peak of Ly$\alpha$ lines in the upper panel reaches about $36.65$.
    }
    \label{fig:composite_spectrum}
\end{figure*}

\subsection{Composite Spectrum of the LAE sample}
\label{subsec:composite_spectrum}
One approach to utilize low signal-to-noise ratio (S/N) individual spectra is to combine them together and obtain high S/N composite spectra. 
This approach has been used in spectroscopic analysis to study both weak emission and absorption spectral features originated from gas in and out of galaxies \citep[e.g.,][]{Zhu2015, Zhang16, Lan2017, Steidel2018, Davis_2023_50kLAE, changdib, Ng2025}. 

Here we focus on the composite spectrum of the LAE sample from which we can extract both weak emission and absorption features associated with ISM and gas flows. We adopt the procedure described in Section~\ref{sub:final_catalog} to obtain a high S/N composite spectrum. Since the number of contributing spectra is smaller at the two ends of the composite spectrum due to the redshift distribution of the LAE sample, the S/N in these regions is lower compared to the central part. Therefore, we trim the edges and retain only the wavelength range from 910 to 3280$\mathrm{\AA}$. In Figure~\ref{fig:composite_spectrum}, we show our DESI LAE composite spectrum in black as well as the HETDEX LAE composite spectrum in grey based on $\sim50,000$ individual spectra from \citet{Davis_2023_50kLAE} for comparison.
In the upper panel, one can see that, besides Ly$\mathrm{\alpha}$ emission, there are many weak metal lines detected in both our DESI composite spectrum and the HETDEX composite spectrum. Given that DESI has wider wavelength coverage (from $3600 - 9800\mathrm{\AA}$) than HETDEX
($3500 - 5500\mathrm{\AA}$) \citep{Hill_HETDEX_2008, HETDEX_2021}, our composite spectrum extends to longer wavelength regions and covers more spectral lines, e.g. weak Fe~\textsc{ii} lines around $2400 \mathrm{\AA}$ and $2610 \mathrm{\AA}$ and Mg~\textsc{ii} $\lambda\lambda2796, 2803$, than the HETDEX composite spectrum. 
The lower panel shows a zoom-in version of the composite spectrum with $0.5 < f_{\lambda} < 1.35$ over $2250\mathrm{\AA} < \lambda_{\text{rest}} < 2850\mathrm{\AA}$, illustrating that weak emission and absorption lines can be clearly detected. 
Moreover, the DESI spectra have a relatively high spectral resolution with $R\sim2000 - 5000$ from 3800 $\rm \AA$ to 9800 $\rm \AA$, capturing the gas kinematics reflected in the line profiles.

The spectral features in Figure~\ref{fig:composite_spectrum} can be used to explore the properties of gas inside or around LAEs.
Here we briefly describe the astrophysical information that can be learned from the spectral lines and defer the detailed investigation in the follow-up study. 
For example, the middle panel of Figure~\ref{fig:composite_spectrum} shows multiple P-Cygni features \citep{Beals_pcygni_1934}, such as N~\textsc{v} $\lambda1241$ and C~\textsc{iv} $\lambda1549$. The presence of P-Cygni profiles indicates strong stellar winds driven by the radiation of young massive stars  \citep[e.g.,][]{Castor_wind_1975, Lamers_wind_1999}. As suggested by \citet{Chisholm_star_2019}, N~\textsc{v} P-Cygni profile implies the existence of massive stars with very young stellar age ($<5\, \mathrm{Myr}$). Similarly, C~\textsc{iv} P-Cygni profile traces stellar wind originating from $<10\, \mathrm{Myr}$. Additionally, the broad He~\textsc{ii} emission as shown in the figure supports the existence of evolved Wolf–Rayet stars with their main-sequence lifetime being shorter than 5 Myr \citep[e.g.][]{Abbott_WRstar_1987, Crowther_WRstar_2007}. Moreover, low-ionization gas in galactic outflows or the interstellar medium (ISM) of star-forming galaxies can be traced by the Fe~\textsc{ii} emission and absorption features between $2300\mathrm{\AA}$ and $2650\mathrm{\AA}$ along with Mg~\textsc{ii} resonant lines $\lambda\lambda2796, 2803$ \citep[e.g.,][]{Rubin_FeII_2011, Prochaska_outlow_model_2011, Zhu2015}.
Furthermore, the Mg~\textsc{II} emission lines shown in the composite spectrum have been considered to be a potential indicator of Ly$\alpha$ and Lyman continuum escape fraction \citep[e.g.,][]{Chisholm_2020, MgII_escape_2022}. The detection of Mg~\textsc{II} emission lines illustrates the potential of utilizing this dataset to explore such connections at z$\sim2-3$.

Finally, our CNN model identifies LAEs and estimates the corresponding redshifts based on the peaks of Ly$\mathrm{\alpha}$ lines when a single line is observed. When there are double-peak Ly$\mathrm{\alpha}$ lines, the redshifts are determined by the peaks of the red component of Ly$\mathrm{\alpha}$ lines. 
We note that the composite spectrum shown in Figure~\ref{fig:composite_spectrum} is based on the peaks of Ly$\mathrm{\alpha}$ lines instead of the systemic redshifts ($z_{\mathrm{sys}}$) of the LAEs.  It is known that the redshift estimation based on Ly$\mathrm{\alpha}$ lines might not entirely reflect the $z_{\mathrm{sys}}$ of the LAEs \citep[e.g.,][]{Trainor_2.7LAE_2015}. Several studies have attempted to estimate the systemic redshift from the Ly$\mathrm{\alpha}$ line profile and other spectral features, using empirical or radiative-transfer–motivated approaches \citep[e.g.,][]{recover_z_2018, recover_z_2019, recover_z_2021, Herrera_2025_Lyaz}. In our case, it is difficult to correct the offset for individual DESI LAE spectra due to the low S/N of other spectral features and the wavelength range outside strong nebular lines, e.g., [O~\textsc{ii}], [O~\textsc{iii}], and H$\alpha$. Therefore, we adopt a simple approach to estimate the overall Ly$\mathrm{\alpha}$ velocity offset by using the spectral lines not significantly affected by stellar wind or resonant scattering in the composite spectrum.
We select He~\textsc{ii} $\lambda1640$, O~\textsc{iii}] $\lambda1666$, and C~\textsc{iii}] $\lambda\lambda1907,1909$. For He~\textsc{ii}, to separate the narrow and broad components, we use two gaussian profiles to fit He~\textsc{ii}. For C~\textsc{iii}], since the doublet is not clearly resolved, we only perform single-gaussian fitting and use $1907.705\mathrm{\AA}$ as the fiducial wavelength. For each line, we compare the peak of the best-fit gaussian to its corresponding rest wavelength and derive Ly$\mathrm{\alpha}$ velocity offset. The results are $\Delta v=-233.26\pm17.93\,\mathrm{km/s}$, $\Delta v=-269.35\pm11.44\,\mathrm{km/s}$, and $\Delta v=-263.58\pm6.15\,\mathrm{km/s}$ based on the fittings on He~\textsc{ii}, O~\textsc{iii}], and C~\textsc{iii}], respectively with 
the systemic redshifts of LAEs being
\begin{equation}
    z_{sys} = z_{Ly\alpha}+\frac{\Delta v}{c}(1+z_{Ly\alpha}).
\end{equation}
These results are consistent with previous measurements being $\sim200-300 \, \mathrm{km/s}$ \citep[e.g.,][]{Erb_galphy_2010, Shibuya_LAE_2014, Trainor_2.7LAE_2015,Davis_2023_50kLAE}. 
{
We note that the above offsets and uncertainties are based on the composite spectrum of all LAE systems. It is possible that subsets of LAEs with different profiles, e.g., double-peak systems \citep[e.g.,][]{recover_z_2018}, would have systematic uncertainties larger than the uncertainties reported above.}

\section{Summary}
In this work, we develop an automatic CNN-based model to detect LAEs hidden in the DESI DR1 dataset. To this end, we investigate the mis-identification patterns of LAEs by the DESI pipeline and use such information to select possible LAE candidates. We then visually inspect those candidates, construct a truth table as a training, validation and test datasets for training our model. We implement techniques to automatically search for hyper-parameters of the architecture. Our final model yields $\sim95\%$ purity and $\sim96\%$ completeness of detecting LAEs in the test dataset. Applying this model to $\sim2$ million DESI spectra, we identify $19,685$ LAEs from redshift 2 to 3.5. We summarize our main findings below:
\begin{enumerate}
    \item Based on the 62 VI-confirmed LAEs in \citet{Lan_2023}, we find that the $\sim37\%$ of the Ly$\mathrm{\alpha}$ emission lines are  ignored by the pipeline, $\sim34\%$ are 
    identified as quasars , $\sim21\%$ are identified as [O~\textsc{ii}] doublet and $\sim8\%$ are identified as other emission lines. 
    
    \item We select LAE candidates based on Ly$\mathrm{\alpha}$ and [O~\textsc{ii}] information and perform visual inspection to build a truth table consisting of $\sim3,200$ LAEs and $\sim3,200$ non-LAEs for training our model.
    
    \item We apply our model to 2,016,328 ELGs with $\Delta \chi^{2}<20$ or o2c $< 0.9$, combine post-processing analysis and construct a LAE sample, consisting of $19,685$ systems. 

    \item Our exploration of the sample shows that the LAEs are blue galaxies at the bright end of Ly$\mathrm{\alpha}$ luminosity function with $L_{\mathrm{Ly\alpha}}>10^{43}\,\rm erg/s$ from redshift 2 to 3.5. The composite spectrum further shows rich spectral information, including the P-Cygni features (N~\textsc{v} and C~\textsc{iv}) along with broad He~\textsc{ii} emission and FeII absorption/emission and MgII emission lines in the longer wavelength that is not covered in previous studies. These observed features inform the physical properties of LAEs as well as the properties of gas flows in and around LAEs. 

    \item  Using non-resonant line information from the composite spectrum, we estimate the average velocity offset ($\sim250$ km/s) between the redshifts based on Ly$\alpha$ emission lines and the systematic redshifts of galaxies.
\end{enumerate}
Our catalog provides a large LAE sample for studying the ensemble properties of high Ly$\mathrm{\alpha}$ luminosity LAEs. Our CNN model along with the catalog can also help the preparation of the DESI-II survey which targets LAEs as a tracer for large-scale structure.
The planned DESI-II survey will observe LAEs with a Ly$\alpha$ flux limit being $\sim10^{-16}$ $\rm erg/s/cm^{2}$ \citep{Schlegel_desi2_2022, Uzsoy2025}, which corresponds to $\sim4\times10^{42}$ erg/s at redshift 2.5, a value similar to our detected LAEs. Therefore, we expect that our model will be applicable to the DESI-II LAE sample. Our model can provide an independent redshift estimation for LAEs, and the LAE catalog can serve as a test dataset for DESI-II pipeline.

We plan to keep improving our model by using the entire spectrum as the model input, enabling our model to reduce false-positive rate using other spectral lines and also allowing our model to recover LAEs at $z > 3.5$. For example, one way is to incorporate attention-based Transformer architectures \citep[e.g., AstroCLIP;][]{AstroCLIP} to capture correlations between multiple emission lines like Ly$\alpha$ and CIV, further enhancing the model capability to distinguish LAEs associated with AGNs.
We will also explore the technique or architecture that can simultaneously provide robust prediction and quantitative uncertainty. 
In addition, we will include more spectra, such as LAEs detected by the DESI-II pilot observations \citep[e.g.,][]{DESI_II_clustering,Raichoor2026_selection} and the ODIN-DESI datasets \citep[e.g.,][]{DESI_II_clustering_w, ODIN_DESI}, in the training sample to enhance the diversity of the Lyman alpha line profiles. We will also train and test our model with sky spectra to better improve and understand the performance of the model in low signal-to-noise regions.
Moreover, we will study the physical properties of LAEs, e.g. light-weighted stellar age, extinction, and metallicity. In addition, from individual spectra, we observe a variety of Ly$\alpha$ emission line profiles which reflect the underlying mechanisms and gas distribution \citep[e.g.][]{Lya_fitting_ML_2022, blueprint, Mukherjee2025}. We plan to explore the diversity of line profiles and their correlations with galaxy properties in a follow up study.
The results will advance our understanding of LAE population not only in the context of galaxy evolution but also for their application as a cosmological probe.

\section*{Data Availability}
The LAE catalog and all data points shown in the figures are available in Zenodo at \dataset[doi: 10.5281/zenodo.22667857]{https://doi.org/10.5281/zenodo.22667857}.

\section*{Acknowledgments}
We thank Ana Sofia Uzsoy and Kyle Dawson for their comments and suggestions for the early version of this paper. JKC and TWL acknowledge supports from National Science and Technology Council (MOST 111-2112-M-002-015-MY3, NSTC 113-2112-M-002-028-MY3), Yushan Fellow Program by the Ministry of Education (MOE) (NTU-110VV007, NTU-111VV007-2, NTU-112VV007-3, NTU-113VV007-4, and NTU-114VV007-5), and MOE-115-YSFMS-0003-001-P2 (NTU 115V2020-1), National Taiwan University research grant (NTU-CC-111L894806, NTU-CC-112L894806, NTU-CC-113L894806) and the summer student program of Academia Sinica, Institute of Astrophysics and Astronomy (ASIAA). SS acknowledges support from the U.S. Department of Energy, Office of Science, Office of High Energy Physics under grant No. DE-SC0024694.

This material is based upon work supported by the U.S. Department of Energy (DOE), Office of Science, Office of High-Energy Physics, under Contract No. DE–AC02–05CH11231, and by the National Energy Research Scientific Computing Center, a DOE Office of Science User Facility under the same contract. Additional support for DESI was provided by the U.S. National Science Foundation (NSF), Division of Astronomical Sciences under Contract No. AST-0950945 to the NSF’s National Optical-Infrared Astronomy Research Laboratory; the Science and Technology Facilities Council of the United Kingdom; the Gordon and Betty Moore Foundation; the Heising-Simons Foundation; the French Alternative Energies and Atomic Energy Commission (CEA); the National Council of Humanities, Science and Technology of Mexico (CONAHCYT); the Ministry of Science, Innovation and Universities of Spain (MICIU/AEI/10.13039/501100011033), and by the DESI Member Institutions: \url{https://www.desi.lbl.gov/collaborating-institutions}. Any opinions, findings, and conclusions or recommendations expressed in this material are those of the author(s) and do not necessarily reflect the views of the U. S. National Science Foundation, the U. S. Department of Energy, or any of the listed funding agencies.

The authors are honored to be permitted to conduct scientific research on I'oligam Du'ag (Kitt Peak), a mountain with particular significance to the Tohono O’odham Nation.

\software{SciPy \citep{2020SciPy-NMeth}, NumPy \citep{harris2020array}, Astropy \citep{astropyI, astropyII, astropyIII}, PyTorch \citep{pytorch}, Torchvision \citep{torchvision}, Ray \citep{liaw_tune_2018}, Optuna \citep{Akiba_optuna_2019}, and Matplotlib \citep{matplotlib}. \\}

\newpage
\appendix
\section{Example Spectra of Visual Inspection and HETDEX outliers}
\label{app:VI_ex}

\begin{figure*}
    \centering
    \includegraphics[width=1\linewidth]{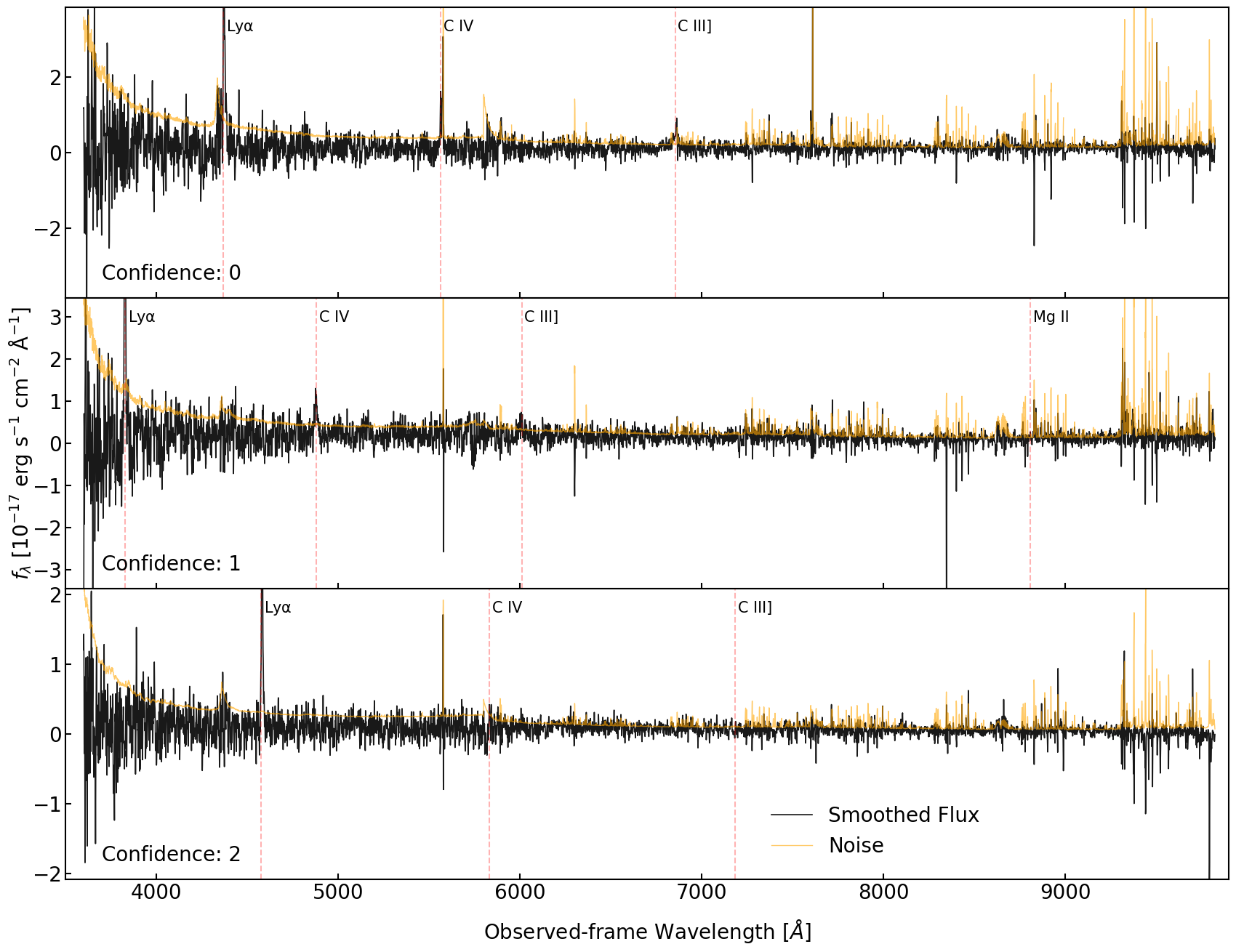}
    \caption{Example spectra of three VI confidence levels. In each panel, we also indicate the major spectral lines based on the best-fit redshifts from Redrock. The upper panel shows an example spectrum with confidence level = 0. This spectrum has significant C~\textsc{iv} and C~\textsc{iii}], indicating that this source is likely a narrow-line AGN instead of a LAE. Most of the sources with confidence level = 1 have weak C~\textsc{iv} line or C~\textsc{iii}]. The spectrum in the middle panel is an example. 
    The spectrum in the lower panel only shows a narrow Ly$\mathrm{\alpha}$ and the confidence level is assigned to 2.}
    \label{fig:VI_ex}
\end{figure*}

\textbf{Example spectra for visual inspection -} 
To construct a true table of LAEs for training our model, we visually inspect spectra of LAE candidates selected using the two methods described in Section~\ref{subsec:selection}. During the visual inspection, each inspected spectrum receives a confidence level, ranging from zero to two. Higher confidence level values imply that the corresponding spectrum is more likely to be a LAE. The values of confidence level are primarily based on the width of Ly$\alpha$ and the apparent strength of the spectral lines other than Ly$\alpha$. In the Figure~\ref{fig:VI_ex}, we show three example spectra with the three confidence levels. The upper panel shows the spectrum with confidence level equal to 0 given the presence of strong C~\textsc{iv} or C~\textsc{iii}]. With the relatively weak detected C~\textsc{iv} and C~\textsc{iii}], the confidence level of the spectrum in the middle panel is assigned as 1. The spectrum in the lower panel only presents narrow Ly$\alpha$ with the confidence level assigned as 2.  In this work, since LAE spectra intrinsically demonstrate weak C~\textsc{iv} and C~\textsc{iii}] signals \citep{Steidel2018,Davis_2023_50kLAE}, we consider sources with confidence level values $\geq$ 1 to be real LAEs. 

\begin{figure*}
    \centering
    \includegraphics[width=1\linewidth]{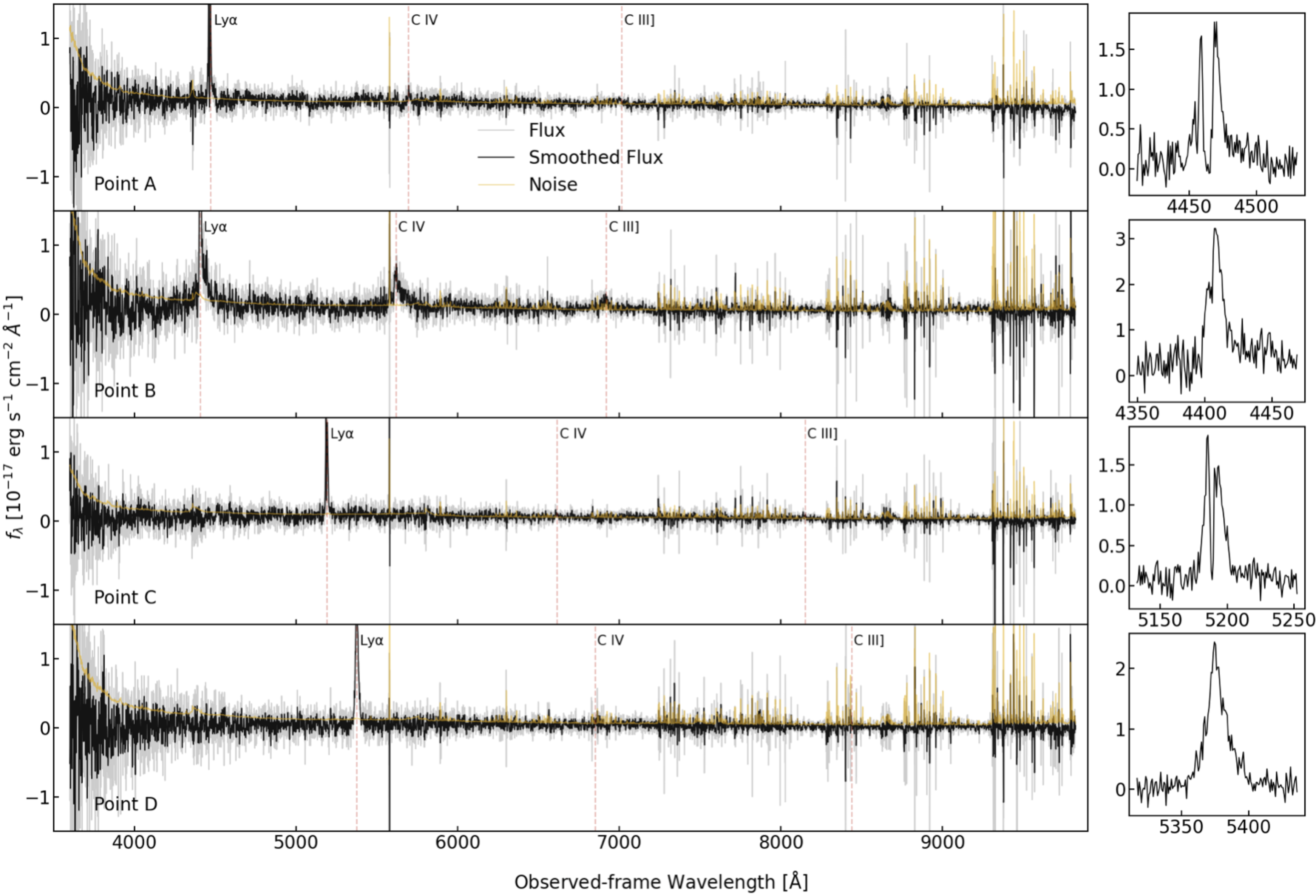}
    \caption{Spectra of the four marked LAEs in Figure 8. Both source A and source C spectra demonstrate double-peak signatures with blue peaks being nearly equal strong (amplitude) as red peaks. Since the training LAEs having Ly$\alpha$ profiles similar to the two here are very rare, it is likely that our model confuse this kind of Ly$\alpha$ with [O~\textsc{ii}]. Source B has strong AGN features (C~\textsc{iv} and C~\textsc{iii}]), and its broad Ly$\alpha$ prevents our model from classifying it as a LAE. Although the last one does not contain significant C~\textsc{iv} or C~\textsc{iii}], its relatively broad Ly$\alpha$ ($\sigma_{v}\sim430\mathrm{km/s}$) also hinders our model from classifying it as a LAE.}
    \label{fig:hetdex_out}
\end{figure*}

\textbf{HETDEX sources with low probabilities -} In Section \ref{sub:DESI-HETDEX}, we tested our model performance with the DESI-HETDEX LAEs and showed that there are 4 sources identified in the sample with low probabilities from our model. In Figure~\ref{fig:hetdex_out}, we show the spectra of the 4 sources. Source A and Source C have strong double-peak line profiles with the blue peaks being as strong as the red peaks. Source B and Source D have broad Ly$\alpha$ lines. The broad Ly$\alpha$ lines are preferentially excluded in our training data and strong double-peak systems are rarely seen in the DESI selected LAEs. Therefore, our model does not predict them as LAEs.

\section{Training Procedures \& Hyper-parameter Tuning}
\label{subsec:training_procedure}
Here we describe the details of the entire training process. 

\textbf{The loss function:} The loss function of our model is a weighted average of two individual loss functions. More specifically, since the second part is to fit the Ly$\mathrm{\alpha}$ emission line with a Gaussian profile, the loss function of our model is defined as
\begin{equation}
    \mathcal{L} = \frac{1}{n}\sum_{i=1}^{n}\mathcal{L}_{i},
\end{equation}

\begin{equation}
    \left \{ 
    \begin{array}{ccc}
    \mathcal{L}_{i} = \frac{1}{1+\alpha}(\mathcal{L}^{\mathrm{BCE}}_{i} + \alpha \, \mathcal{L}^{\mathrm{MSE}}_{i}), & \textrm{if} & \mathrm{real \, LAE} \\
    \mathcal{L}_{i} = \mathcal{L}^{\mathrm{BCE}}_{i}, & & \mathrm{otherwise} \\
    \end{array}
    \right.
\end{equation}
where $n$ is the batch size, $\mathcal{L}_{i}$ is the loss for the $i^{\mathrm{th}}$ spectrum in a certain batch, $\mathcal{L}^{\mathrm{BCE}}_{i}$ is binary cross entropy (\texttt{BCELoss}) and is the loss for the first part output, $\mathcal{L}^{\mathrm{MSE}}_{i}$ is mean squared error (\texttt{MSELoss}) and is the loss for the second part output, and $\alpha$ is the loss weight. Note that the batch size and the loss weight are tunable hyper-parameters, and the total loss $\mathcal{L}$ is available for calculating gradient descent and backward propagation. According to the loss functions defined above, the loss for the second part of our model is considered only if the input spectrum is a real LAE (based on the VI described in the section~\ref{subsec:selection}). 
The binary cross entropy used in this work is defined as
\begin{equation}
    \mathcal{L}^{\mathrm{BCE}}_{i} = -\sum_{j=1}^{m_{1}} y_{ij,1}\ln x_{ij,1} + (1 - y_{ij,1})\ln(1-x_{ij,1}),
\end{equation}
where $m_{1}$ is the number of pixels in the output array of the first part of our model, which is 76, $x_{ij,1}$ is the predicted probability at the $j^{\mathrm{th}}$ pixel in the $i^{\mathrm{th}}$ output array within a certain batch, and $y_{ij,1}$ is the corresponding ground-truth label. Similarly, the mean squared error here is defined as
\begin{equation}
    \mathcal{L}^{\mathrm{MSE}}_{i} = \sum_{j=1}^{m_{2}}(x_{ij,2} - y_{ij,2})^2,
\end{equation}
where $m_{2}$ is the number of pixels in the output array of the second part of our model, which is 97, $x_{ij,2}$ is the model prediction at the $j^{\mathrm{th}}$ pixel in the $i^{\mathrm{th}}$ output array within a certain batch, and $y_{ij,2}$ is the corresponding ground-truth label. During the training stage, we provide a correct probability array and a Gaussian at the correct location and width to train the model. \\\\

\textbf{Automatic search for the hyper-parameters:} In our model, there are 12 trainable hyper-parameters which are summarized in Table~\ref{tab:hyperparameter}.  
The sampling space of the hyper-parameters is too large to perform grid-search for each of the points. To address the problem, we divide the training process into four stages with each, focusing on a different subset of hyper-parameters and fixing the rest of the hyper-parameters. In the first stage, we focus only on global hyper-parameters, including the learning rate, batch size, and loss weight. In the next stage, we update and fix the hyper-parameters tuned in the first stage and focus on the hyper-parameters more relevant to the first part of the model, including kernel size $1.1\sim 1.4$, and channel factor --- a factor multiplied to the number of convolution channels in the first part. In the third stage, we also update the hyper-parameters explored in the second stage and tune kernel size 2.1, 2.2, fc layer width, and dropout rate. In the final stage, we tune all the hyper-parameters again but with much smaller sampling space to fine-tune the model and find the best-fit hyper-parameters (shown in the rightmost column of Table \ref{tab:hyperparameter}).

During the stage-wise hyper-parameter tuning, we utilize two packages, Ray and Optuna \citep{liaw_tune_2018, Akiba_optuna_2019}, to facilitate automated and efficient exploration of the large sampling space, as well as to record the outcomes of each trial and epoch. To allocate the limited GPU computing resources, we adopt Asynchronous Successive Halving Algorithm (ASHA) \citep{Li_ASHA_2018}, a commonly used and powerful early-stopping algorithm for dealing with large sampling space. Basically, ASHA trains multiple candidate trials in parallel and periodically evaluates their intermediate performance. The trials that have poor performance are stopped early. Only the trials with better performance are allowed to continue training with increased computational budget. In this way, computational resources are efficiently allocated to promising configurations of hyper-parameters. In practice, we use \texttt{ASHAScheduler} from Ray package, with three primary parameters, including \texttt{max\_t}, \texttt{grace\_period}, and \texttt{reduction\_factor}. The parameter \texttt{max\_t} sets the maximum training epochs for each trial, \texttt{grace\_period} specifies the minimum training epochs before a trial is evaluated, and \texttt{reduction\_factor} controls the fraction of trials that are retained at each evaluation point. In our case, with \texttt{max\_t = 120}, \texttt{grace\_period = 40}, and \texttt{reduction\_factor = 2}, the scheduler defines three levels by epoch range: level 1 (epoch = $1\sim40$), level 2 (epoch = $41\sim80$), and level 3 (epoch = $81\sim120$). Then, every time a trial within a level is finished, the trials with performances in the top $50\%$ of each level will be promoted to the next level. If there is no trial that can be promoted to the next level, a new trial (new hyper-parameter configuration) will be started in the bottom level (level 1).

An effective hyper-parameter sampler is crucial for efficiently exploring a large and high-dimensional hyper-parameter space, as it determines how new configuration candidates are selected based on the results of previous trials. The hyper-parameter sampler used in this work is the \texttt{TPESampler} from Optuna package. The basic concept of this sampler is that this method models two probability distributions of hyper-parameters based on the previous trials: one derived from the configuration with better performances, denoted as $l(x)$, and the other corresponding to the configurations with poorer performances, denoted as $g(x)$. New hyper-parameter configuration candidates are sampled by prioritizing regions of hyper-parameter space with higher values of $l(x)/g(x)$, increasing the likelihood of selecting promising configurations in the subsequent trials. For more details about the principle of the sampler, we refer the readers to \citet{NIPS2011_86e8f7ab}.

\textbf{Best-fit hyper-parameters:} To determine the hyper-parameters with the best performance, we use three metrics, including accuracy of the first part --- defined as the number of correct classifications divided by the total number of spectra, $\mathcal{L}_{\mathrm{BCE}}$, and $\mathcal{L}_{\mathrm{MSE}}$. To reduce the impact of short-term performance fluctuations on the hyper-parameter selection during the training process, for each trial, we select the epochs with top $95\%$ accuracy performance and use the median values of accuracy, $\mathcal{L}_{\mathrm{BCE}}$, and $\mathcal{L}_{\mathrm{MSE}}$ of those epochs to represent the performance of the trial. 
To simultaneously account for the three metrics mentioned above, we adopt the concept of the Pareto frontier \citep{pareto_cours_1896, Deb_NSGA_2002}, which is commonly used in the field of multi-objective optimization. The key concept is that a trial is considered to lie on the Pareto frontier if no other trial demonstrates better performance in all three metrics at the same time (i.e., higher accuracy, lower $\mathcal{L}_{\mathrm{BCE}}$, and lower $\mathcal{L}_{\mathrm{MSE}}$). In each of four stages, we apply this Pareto frontier analysis on all of the trials regardless of their terminated epochs. Usually there are multiple trials on each Pareto frontier. For simplicity, we select the one with the lowest sum of $\mathcal{L}_{\mathrm{BCE}}$ and $\mathcal{L}_{\mathrm{MSE}}$, and its hyper-parameter configuration will be fixed in the next stage. In the fourth stage (last stage), among 280 trials there are 16 trials on the Pareto frontier, and the trial with the lowest sum of $\mathcal{L}_{\mathrm{BCE}}$ and $\mathcal{L}_{\mathrm{MSE}}$ is considered our best-fit hyper-parameter configuration, which is listed in the rightmost column of Table \ref{tab:hyperparameter}. We note that during this hyper-parameter optimization phase, we use the training dataset to train the model and use the validation dataset to estimate the performance metrics. 

Finally, with the final optimized hyper-parameters, we combine the training and validation datasets to train the corresponding weights of the model, and evaluate its performance on the test dataset. This model serves as our finalized automatic LAE identification algorithm.

\begin{table*}
    \centering
    \caption{Tunable Hyper-parameters of the Model}
    \begin{tabular}{llcc}
        \hline
        \hline
        \textbf{Hyper-parameter} & \textbf{Description} & \textbf{Sampling Space} & \textbf{Best-fit} \\
        \hline
        Learning rate & Learning rate for Adam optimizer & ($10^{-5}, 10^{-1}$, float) & 0.0011 \\
        \hline
        Batch size & Number of spectra for each batch & ($2^3, 2^4, \dots, 2^{10}$) & 64 \\
        \hline
        Loss weight ($\alpha$) & Weight for the weight-averaged loss function & ($10^{-6}, 10^{6}$, float) & 4.5 \\
        \hline
        Kernel size 1.1 & Kernel size for the 1st convolution layer in the first part & (15, 17, $\dots$, 37) & 25 \\
        \hline
        Kernel size 1.2 - 1.4 & Kernel size for the 2nd - 4th convolution layer in the first part & (3, 5, $\dots$, 15) & 9, 7, 7 \\
        \hline
        Channel factor & Factor multiplied to the number of convolution channels in the first part & ($2, 2^{2}, \dots, 2^{8}$) & 64 \\
        \hline
        Kernel size 2.1, 2.2 & Kernel size for the 1st, 2nd convolution layer in the second part & (3, 5, $\dots$, 15) & 7, 5\\
        \hline
        Fc layer width & Width of the fully connection layers in the second part & ($2^6, 2^7, \dots, 2^{10}$) & 512 \\
        \hline
        Dropout (p) & Dropout rate & (0, 0.1, \dots, 0.5) & 0.2 \\
        \hline
    \end{tabular}
    \label{tab:hyperparameter}
\end{table*}

\section{Results of the AGN Sample}
\label{subsec:LAE-AGN}
Here we provide the results of the AGN sample. We note that the completeness and purity of this sample are not estimated due to the fact that our training, validation, and test datasets focus on LAEs powered by star formation activities.

\begin{figure}[ht!]
    \centering
    \includegraphics[width=0.6\linewidth]{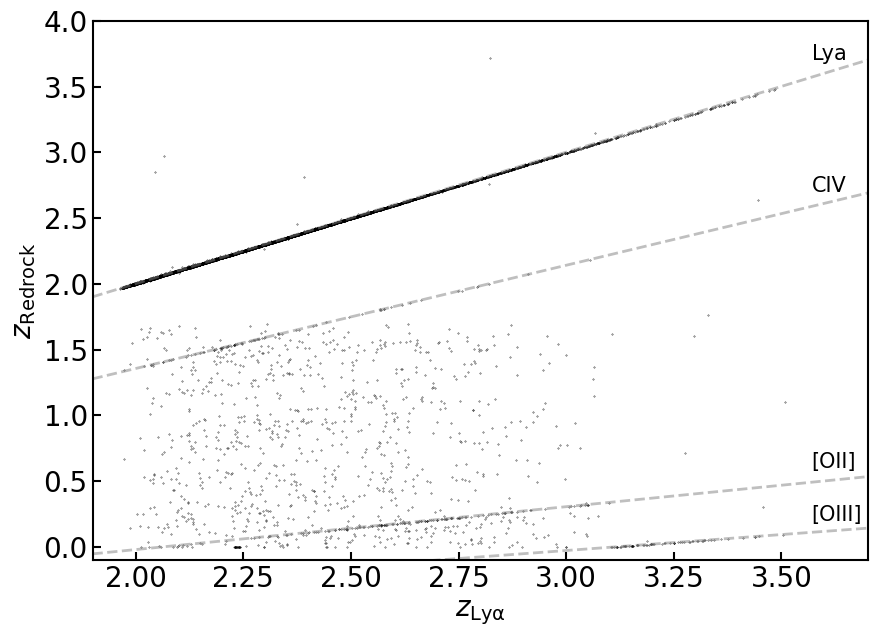}
    \caption{Distribution of the AGN redshifts predicted by our model ($z_{\mathrm{Ly\alpha}}$) and Redrock ($z_{\mathrm{Redrock}}$).
    Several straight lines are highlighted using gray dashed lines, and the spectral lines to which Ly$\alpha$ is identified are labeled on the right.}
    \label{fig:zm_zr_AGN}
\end{figure}

\textbf{Redshift estimations:} Figure~\ref{fig:zm_zr_AGN} shows the Ly$\alpha$ redshifts from our model (x-axis) and the Redrock redshifts (y-axis). The pattern is similar to the pattern for the LAE sample shown in Figure~\ref{fig:zm_zr}. The main difference is that the AGN sample has (1) a lower fraction of sources with $z_{\mathrm{Redrock}} < 1.7$, and (2) a higher fraction of sources with consistent redshift estimations from the two methods. There are $80.6\%$ of sources with consistent redshifts from Redrock and our model, $1.0\%$ of Ly$\alpha$ lines misidentified as C~\textsc{iv}, $1.3\%$ of Ly$\alpha$ lines misidentified as [O~\textsc{ii}], and $0.8\%$ of Ly$\alpha$ lines misidentified as  [O~\textsc{iii}]. Most of the remaining $16.3\%$ are sources without detecting Ly$\alpha$ lines by Redrock. We note that for the LAE sample, only $17.8\%$ of sources with consistent redshifts from Redrock and our model, indicating that the sources at higher C~\textsc{iv}/Ly$\alpha$ flux ratio tend to be identified as quasars with correct redshifts by Redrock.

\begin{figure}[htbp]
  \centering
  \begin{minipage}{0.48\textwidth}
    \centering
    \includegraphics[width=\linewidth]{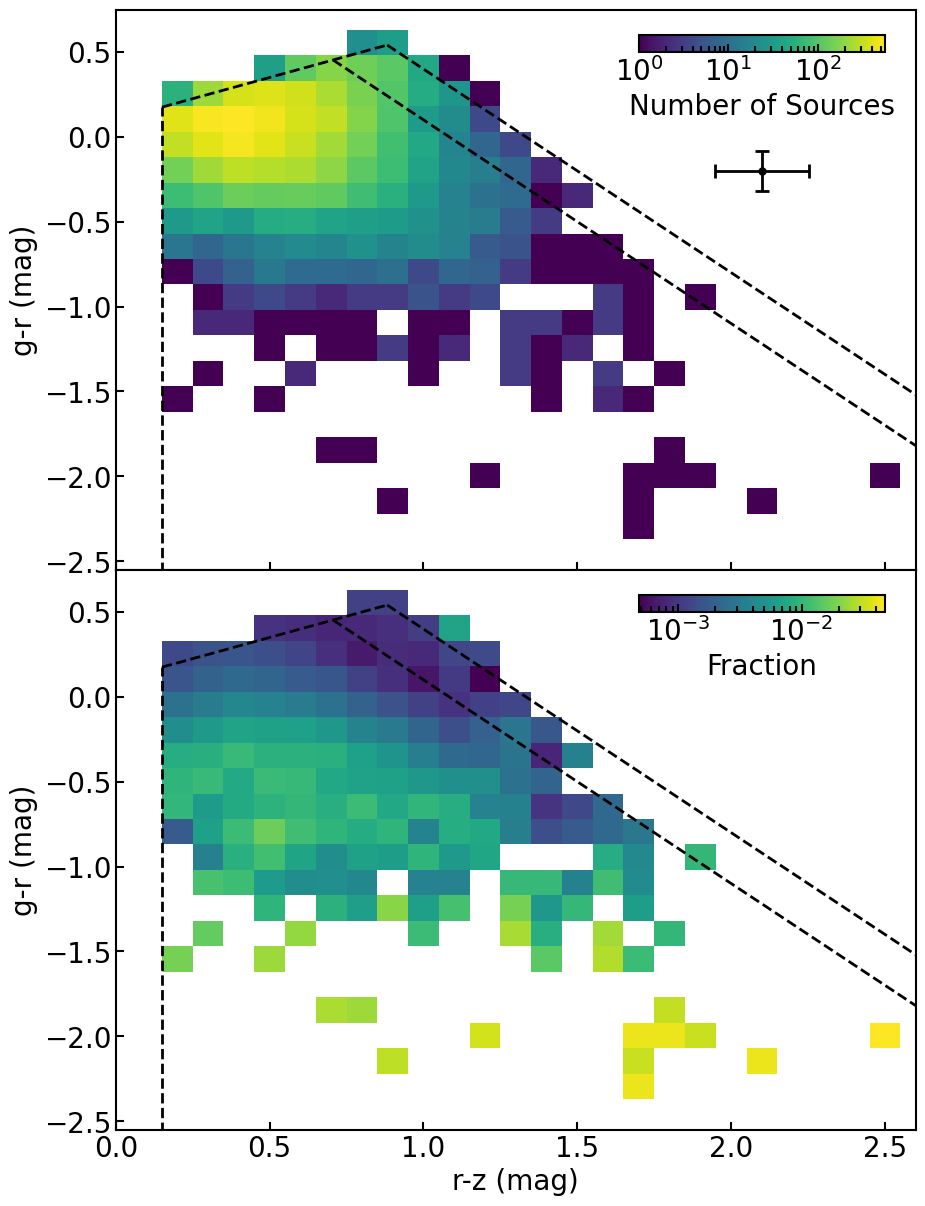}
    \caption{Upper panel: Number of AGNs as a function of g-r and r-z. Lower panel: Ratio of number of AGNs to total number of ELGs in DESI DR1 main survey as a function of g-r and r-z. The typical $1 \sigma$ uncertainties of g-r and r-z are 0.116 and 0.154 mag, respectively, as shown by the error bar.}
    \label{fig:color_fraction_AGN}
  \end{minipage}
  \hfill
  \begin{minipage}{0.48\textwidth}
    \centering
    \includegraphics[width=\linewidth]{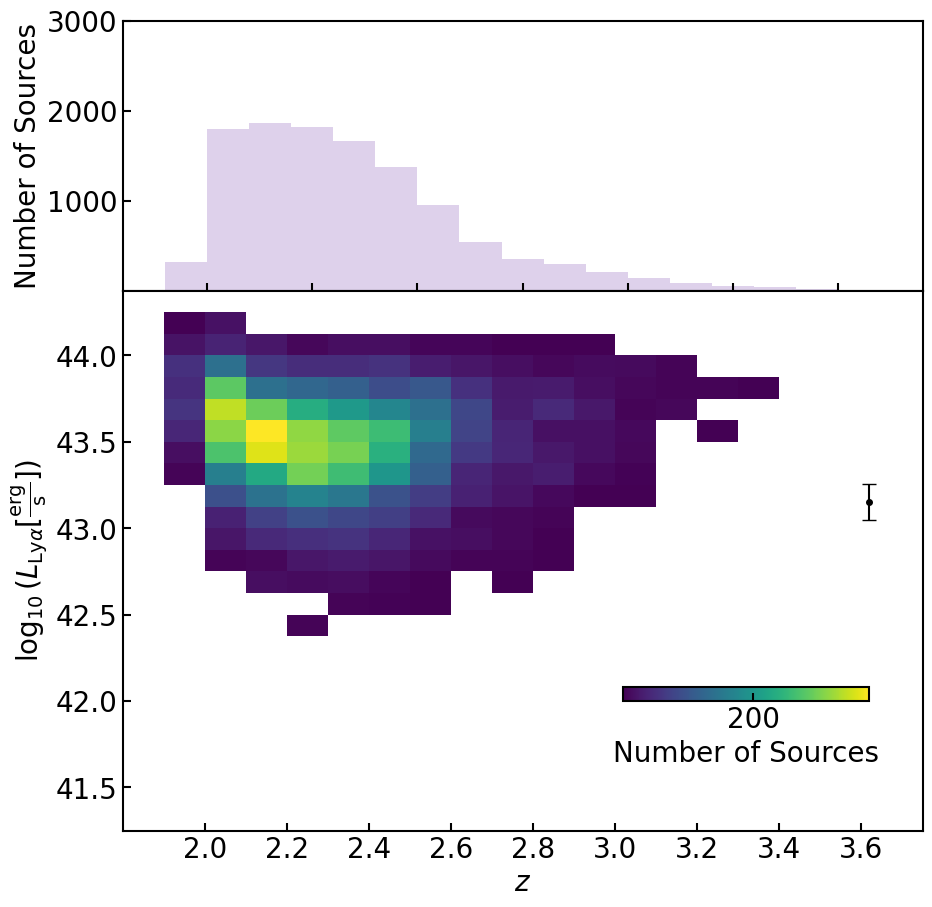}
    \caption{Redshift distribution and Ly$\mathrm{\alpha}$ luminosity as a function of redshift.
    The color code in the lower panel indicates the number of AGNs within the bin. The error bar  demonstrates the typical $1\sigma$ uncertainty (0.105 dex).}
    \label{fig:luminosity_redshift_AGN}
  \end{minipage}
\end{figure}

\textbf{Color distributions:} The upper panel of Figure~\ref{fig:color_fraction_AGN} shows the number of AGNs in the $g-r$ and $r-z$ color space. The lower panel shows the fraction of AGNs in the color-color space. We find that the LAE sample and the AGN sample are concentrated at different color regions with the AGN sample having redder r-z colors than the LAE sample. This reflects different intrinsic physical properties among the two samples.

\textbf{Redshift and Ly$\alpha$ Luminosity:} Figure~\ref{fig:luminosity_redshift_AGN} shows the redshift and Ly$\alpha$ luminosity distributions of the AGN sample. We find that the median Ly$\alpha$ luminosity of the AGN sample is higher than that of the LAE sample by 0.26 dex and the redshift distributions of the two sample are different as well. 

\begin{figure}[ht!]
    \centering
    \includegraphics[width=\linewidth]{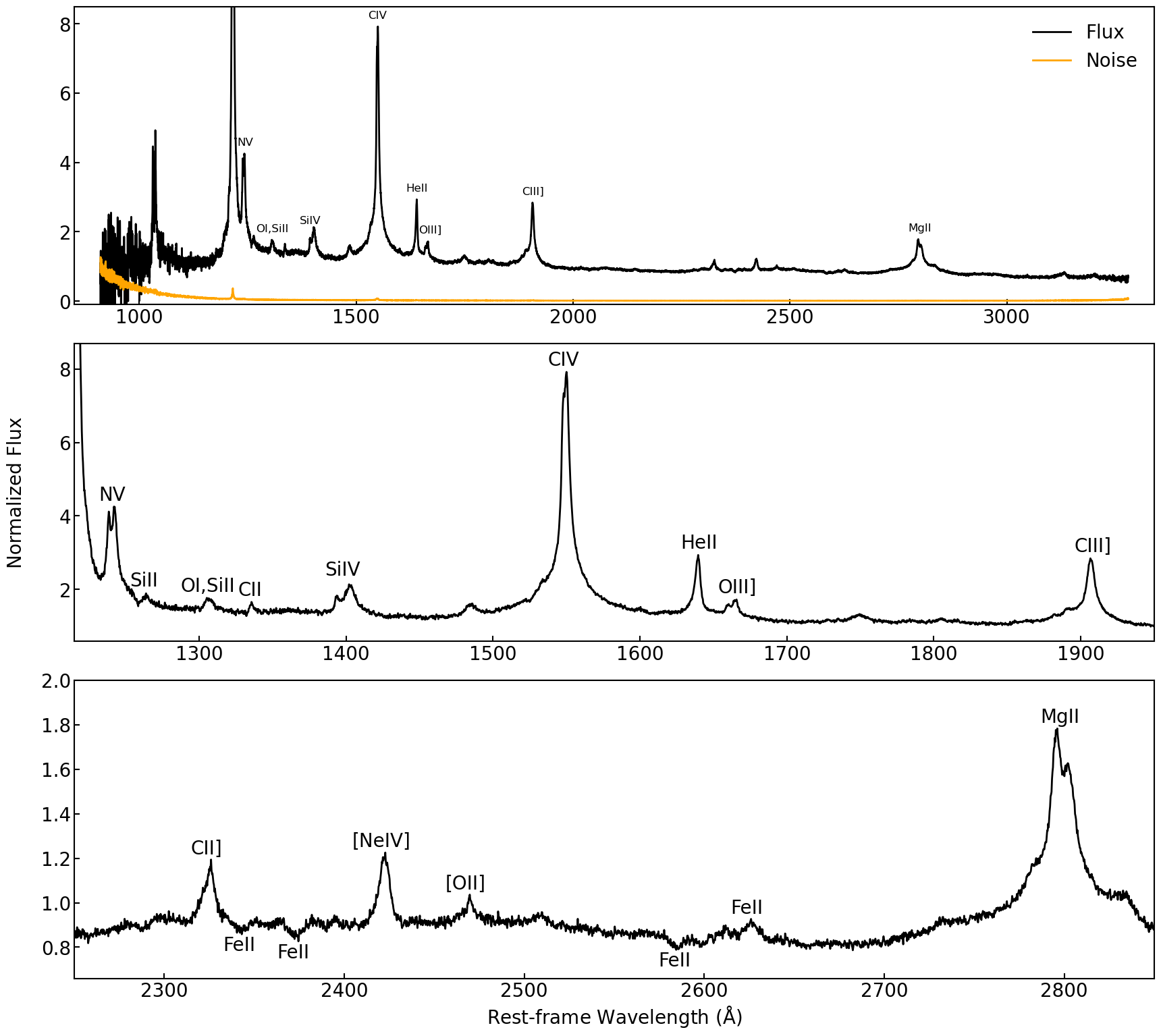}
    \caption{Composite spectrum of 11,505 AGNs. The black and orange curves are the normalized flux of the composite spectrum and the uncertainty estimated from 500 bootstrap realizations, respectively. The middle panel is a zoom-in version of the upper panel, focusing on $1215 < \lambda_{\text{rest}} < 1950 \mathrm{\AA}$. The lower panel is also a zoom-in version of the upper panel, focusing on $2250 < \lambda_{\text{rest}} < 2850 \mathrm{\AA}$. We note that the peak of Ly$\alpha$ lines in the upper panel reaches about 44.06.}
    \label{fig:composite_spectrum_AGN}
\end{figure}

\textbf{Composite spectrum of the AGN sample:} Figure~\ref{fig:composite_spectrum_AGN} shows the composite spectrum for the AGN sample. Compared to the composite spectrum of the LAE sample in Figure~\ref{fig:composite_spectrum}, one can tell that there are numerous differences between the two. For example, in the LAE composite spectrum, there are several (P-Cygni-like) features containing blueshifted absorption lines and redshifted emission, e.g., Si~\textsc{ii}, C~\textsc{ii}, and C~\textsc{iv}, while those spectral lines become dominated by emission in the AGN sample. Besides, the profiles of C~\textsc{iii}] and Mg~\textsc{ii} are much broader in the AGN spectrum. All of the properties indicate that Ly$\alpha$ emission of the LAE and AGN samples are powered by distinct physical mechanisms and energy sources.

\bibliography{citation-bibtex}{}

\end{document}